\documentclass{article}
\usepackage{graphicx} 
\usepackage[margin=2cm]{geometry}
\usepackage{subcaption}
\usepackage{algorithm}
\usepackage{algpseudocode}
\usepackage{amsmath}

\title{Timescale-agnostic characterisation for collective attention events}
\author{Tristan J.B. Cann, Iain S. Weaver and Hywel T.P. Williams}
\date{}

\begin{document}

\maketitle

\begin{abstract}
    \noindent Online communications, and in particular social media, are a key component of how society interacts with and promotes content online. Collective attention on such content can vary wildly. The majority of breaking topics quickly fade into obscurity after only a handful of interactions, while the possibility exists for content to ``go viral'', seeing sustained interaction by large audiences over long periods. In this paper we investigate the mechanisms behind such events and introduce a new representation that enables direct comparison of events over diverse time and volume scales. We find four characteristic behaviours in the usage of hashtags on Twitter that are indicative of different patterns of attention to topics. We go on to develop an agent-based model for generating collective attention events to test the factors affecting emergence of these phenomena. This model can reproduce the characteristic behaviours seen in the Twitter dataset using a small set of parameters, and reveal that three of these behaviours instead represent a continuum determined by model parameters rather than discrete categories. These insights suggest that collective attention in social systems develops in line with a set of universal principles independent of effects inherent to system scale, and the techniques we introduce here present a valuable opportunity to infer the possible mechanisms of attention flow in online communications.
\end{abstract}

\section{Introduction}

Following an explosion of popularity over the last two decades, online platforms, and particularly social media services, have become important components of social relationships and media dissemination. Measuring the activity of individual users has been relatively straightforward due to the wealth of data available through social media platforms, but collective attention to a particular topic, event or piece of media remains more difficult to observe, due in-part to the lack of a formal definition. Wu and Huberman~\cite{Wu07} consider collective attention to be the process of how attention to novel topics propagates and eventually fades among large populations. This definition does not consider the source of the attention and whether it is meaningfully ``collective'', that is to say driven by interactions between individuals. Under Wu and Huberman's definition, a large group of individuals simultaneously, but independently, paying attention would be considered collective attention, despite the ``collective'' component influencing nothing more than the number of members in the system. Lin et al.~\cite{Lin14} propose shared attention as ``a temporary state in which the individual members of an audience for an event are mutually aware of each other’s attention to the event'', a definition which incorporates such a requirement. Given the difficulty in measuring this second definition, it is likely that the debate over a formal definition will continue. For the purposes of this manuscript, we impose our own definition of collective attention that draws themes from both of these existing definitions. We define collective attention as the state wherein the attention of others increases an individual's attention to a topic. In this way, we are incorporating a measure of preferential attachment among attention as discussed in the context of network formation by Barab\'asi and Albert~\cite{Barabasi99}, and explicitly require a level of interaction between participants in the system.

Each of these definitions of collective attention alludes to the mechanisms defining how it emerges and evolves over time. Understanding these mechanisms is of central importance to those who seek widespread engagement with an issue, such as governments, journalists or advertisers. Efforts in this direction may provide valuable insights into universal trends in human behaviour and help to distinguish them from platform effects (i.e. those informed by the medium within which the interaction takes place). Such efforts are complicated by the divergent scales across social systems afforded by online social networks, making it easier to create millions of impressions compared to offline settings. Despite these differences, patterns driven by universal communication behaviours should appear in both instances.

In this manuscript we present a novel methodology for analysing activity around peaks of attention across the wide range of time and volume scales observed. This new method addresses two significant shortcomings of existing timeseries approaches and presents a valuable capability for comparing apparently dissimilar phenomena to reveal similar generating mechanisms manifesting at different scales. Primarily our methodological improvement removes the need to determine a suitable timescale \emph{a priori} by measuring the rate of arrival of a portion of the event tweets relative to the length of the interval, alongside a normalisation process to resolve variation in activity volumes. Using this new methodology we analyse a dataset of hashtags used in the Twitter conversation around Brexit, the common term for the United Kingdom leaving the European Union. We demonstrate the effectiveness of our methodology for comparing events at different scales and find that three of the four classes characterised in previous work by Lehmann et al.~\cite{Lehmann12} are recovered, but that their ``peak only'' class is likely an artefact of the resolution limit induced by coarse binning of the timeseries. Instead we find a fourth class where activity patterns abruptly shift between two modes, revealing a new behaviour in hashtag usage that is not well-represented in existing methods.

To support this methodological innovation we define an agent-based model to simulate the usage of topics in social networks. Our modelling framework allows for free variation of a small set of parameters to simulate different behavioural patterns. The key component of our model is the consideration of both local (driven by social contacts) and global attention (driven by the inherent importance of, and interest in, the topic over time) and how these behaviours affect the observed trends over time. By varying the relative importance of these two sources and the temporal patterns of global attention we are able to reproduce the types of behaviour seen in the Twitter data.

The manuscript begins by discussing a range of related work and other attempts to address the challenges with studying collective attention in section~\ref{sec:Background}. Section~\ref{sec:representation} provides a full description of our new technique to create scale-independent representations of collective attention, describe the dataset of Brexit-related tweets and discuss the behavioural patterns we observe. In Section~\ref{sec:model} we complement these results with our agent-based model and link the simulated behaviours from this model to those observed in the Twitter data. We conclude in Section~\ref{sec:discussion} by discussing the consequences of our results, their implications for collective attention in social systems and provide recommendations for how our proposed methods could be applied in future work.

\section{Background and literature review}\label{sec:Background}

There is in extensive suite of methods and tools to analyse timeseries data and study periods of increased attention, such as election debates~\cite{Lin13} and reactions to online news stories~\cite{Castillo14}. Previous work using them to measure collective activity has examined temporal patterns and decay rates of interactions on the news aggregator website Digg~\cite{Wu07}. By measuring the rate of decay in Digg interactions with a number of stories over time, Wu and Huberman were able to demonstrate it followed a pattern of stretched exponential relaxation with a half-life of approximately one hour. Other approaches apply user metadata to predict adoption of hashtags (user-defined tokens signifying particular topical relevance)~\cite{Yang12} or content metadata to identify potentially malicious attempts to spam the online conversation~\cite{Lee12}. Many studies use measures of departure from some baseline to understand systemic change in social networks with some success at detecting and quantifying collective attention events~\cite{Sasahara13} and capturing the evolving phases of attention under some external stimulus~\cite{tenThij19}. In addition to considering only macro-level variations, Lin et al.~\cite{Lin14} and De Domenico and Altmann~\cite{DeDomenico20} show that different types of activity such as retweeting, replying and sharing on the social media platform Twitter respond differently to increased collective attention.

Other attempts to measure collective behaviour make use of the macro-scale tools of network science. Social network structure influences the content users are exposed to, and often include recommender systems which suggest connecting with other users or content based on their immediate neighborhood. For example, by recording the followership of users, it is possible to enhance the prediction of popular memes~\cite{Weng14}. He and Lin~\cite{He17} study the social media platform Twitter by building networks of successive hashtag usage and find that the topology of the underlying network changes around disaster events. Another common approach considers the connectivity between users and other objects in a bipartite network to track how users engage with content over time~\cite{Cann21} and how topical clusters evolve~\cite{Saito15,LorenzSpreen18}. Such studies provide insights into how different types of content respond to increased attention, for example demonstrating that negative aspects such as controversy~\cite{Garimella17} can become more prevalent in such circumstances. Network methods have also been shown to be useful for event detection purposes~\cite{Moutidis19}.

Beyond the techniques used to measure collective attention online, it is common for studies to propose different modelling approaches to investigate the factors that influence the growth of attention at scale. Given the similarities between information diffusion and epidemic spreading, adaptations of the well-known SIR (susceptible-infected-recovered) and SIS (susceptible-infected-susceptible) models for collective attention~\cite{Kitsak10} have been applied in addition to models inspired by cellular growth~\cite{Wang19}. Agent-based models are frequently applied to test the importance of user-level effects such as neighbour behaviour~\cite{Gleeson14} or memory of recently seen content~\cite{Bathina17}. Some models consider network growth in discussions (e.g.~\cite{Medvedev19}) through extensions of the concept of preferential attachment proposed by Barab\'{a}si and Albert~\cite{Barabasi99}. Others still use a variety of mathematical modelling frameworks to capture how information and attention spread, such as: Markov chains~\cite{deArruda17, Akbarpour18}; differential equation models of information flow~\cite{Harush17}; competition between topics for the same limited attention~\cite{Weng12}; and size and lifetime of retweet cascades~\cite{Burnap14}. It has been shown that such models do not need to be complex to mimic observed behaviour at the system-level. Huynh et al.~\cite{Huynh15}, for example, developed a model using only two parameters: virality of the content, and speed of decline after the peak, and are still able to capture a range of observed phenomena from a real-world dataset.

Understanding the processes underlying collective attention presents some valuable opportunities to researchers for understanding the rising threat of misinformation in modern society. Vosoughi et al.~\cite{Vosoughi18} compare the cascade statistics of real and fake news stories, finding that fake news stories spread further, faster and to more unique users than real news stories. Zhao et al.~\cite{Zhao20} complement this by demonstrating that the spreading topologies of real and fake news are characteristically different. On the other hand, Mitra et al.~\cite{Mitra17} exploit these differences to show that veracity classes can be identified by information activity over time. In contrast to their use for problematic behaviours, social networks also provide fertile ground for organising social action and protests and have already been studied around the Arab Spring~\cite{Starbird12}, political rallies~\cite{Ertugrul19} and climate change protests~\cite{Segerberg11}. Such advances will be invaluable for tackling issues arising in a connected society, but ethical consideration needs to be given to ensure that these tools do not adversely affect the rights of social media users and the wider public.

Beyond these attempts to measure and model attention, some effort has been devoted to classifying different attention patterns. The work of Lehmann et al.~\cite{Lehmann12} is often cited when considering the temporal trends of online activity. They studied the distribution of activity around the day of peak attention and found that events fell into four classes that capture different types of behaviour (see Fig.~\ref{sh:fig:eg_classes} for illustrations of the patterns typical of each class). The first class captures events with a gradual build up of attention until the peak before quickly declining. The second class reverses this trend, with a sharp increase in activity at the peak and a gradual decline afterwards. The third class incorporates both a gradual increase and a gradual decrease in attention around the peak and the fourth class covers events in which the activity is concentrated almost entirely on the peak day. Extensions to this work have been published over the years, such as~\cite{Huynh15} which supports the existence of these distinct categories through a generative model. 

These categorisations are useful for understanding system dynamics but make a potentially erroneous assumption by assigning a timescale to the process. Lehmann et al. bin observed tweet counts into one-day intervals to form a timeseries, meaning that events occurring within a single day are represented by a single value, regardless of their underlying behaviour. This may be particularly problematic as observation of social media suggests that different topics can be the focus of attention for anything from minutes (e.g. an unsuccessful attempt at a new viral trend) to weeks (e.g. elections). This variation means that the choice of bin size needs to be carefully considered; daily or hourly aggregation may be the most obvious choices but should be analysed with the caveat that certain features may be over- or under-exposed at the chosen resolution. Furthermore, certain timescales obscure the day-night or weekly cycles, important patterns in social media activity which may need special attention. 

As a more simplistic measure of collective attention, increased activity volume can be seen as increased attention to a topic. What is not clear, however, is whether increased activity is a sign of individual attention growing independently under exogenous factors or collective attention driven by the interactions between users. Such a distinction is often impossible to measure using the data available to scholars of online social systems and as a result peaks of activity are often used as proxies for collective attention. Hashtags have frequently been used to identify such activity peaks since they provide a convenient indicator that the user considers their post relevant to the topic.


After reviewing the previous efforts to quantify collective attention in the literature, we identify three problems that need to be resolved, and use them to define the goals for our proposed methodology. Firstly, the process of timeseries binning can over- or under-expose aspects of event behaviour depending on their inherent timescales. Secondly, directly comparing distinct events is hampered by different volume scales and unknown factors affecting baseline activity rates. Finally, the processes driving the evolution of events are complex or unobservable but perceived behaviour types may be linked to certain process types. These limitations make it difficult to efficiently work with large datasets of collective attention events as while it is easy to aggregate data at the same resolution, it is difficult to know whether the resolution is appropriate for all events considered. Facilitating direct comparisons on the other hand requires significant effort to adopt a scale-independent approach and normalise both the time and volume scales, as well as address any independent factors affecting activity volumes. Finally, the balance of mechanistic factors that influence how social systems propagate information will make definitive attribution difficult, but robust analyses should suggest potential causal mechanisms.

\section{Measuring the shapes of attention}\label{sec:representation}

We now develop a new scale-independent representation for activity in a social system. Through this representation we aim to reflect underlying mechanistic patterns in the processes driving this activity regardless of the size of the social system or the duration of observation. In this section we motivate and explain our developments before testing our new representation on a dataset of Twitter discussions and compare the observed trends to Lehmann et al.'s four dynamical classes~\cite{Lehmann12}. 

\subsection{Defining a robust event representation under scale variation}

\begin{figure}%
    \centering
    \begin{subfigure}{0.45\textwidth}
    \includegraphics[width=\linewidth]{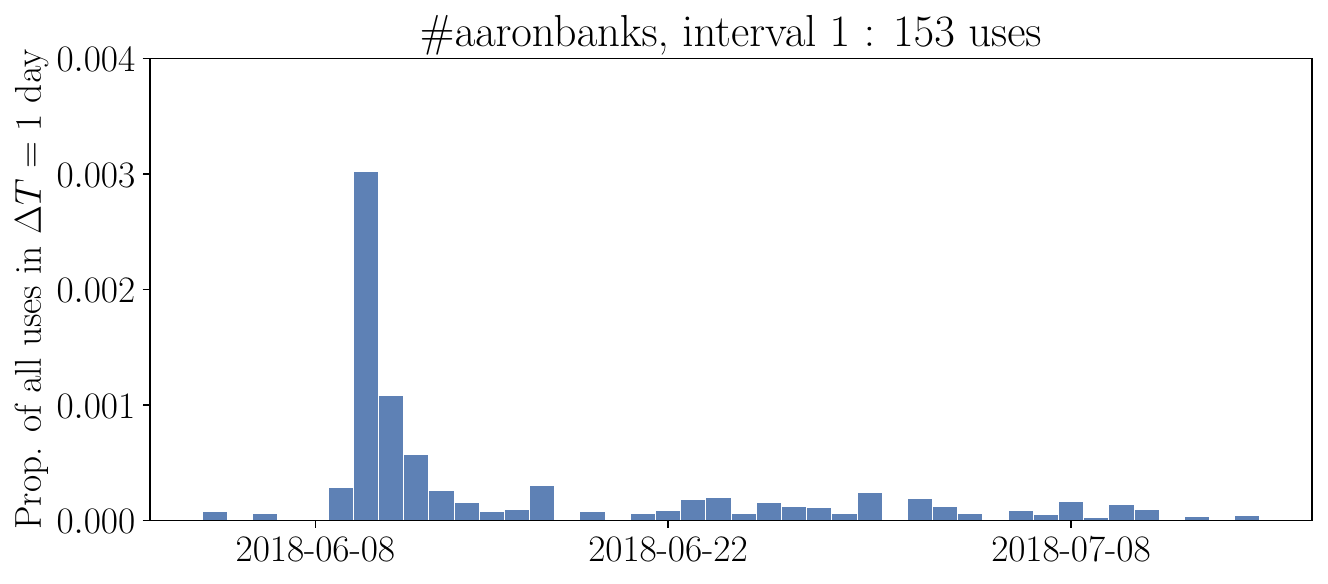}
    \caption{Short-lived peak in days.}\label{sh:fig:aaronbanks_1_day}
    \end{subfigure} 
    \hfill
    \begin{subfigure}{0.45\textwidth}
    \includegraphics[width=\linewidth]{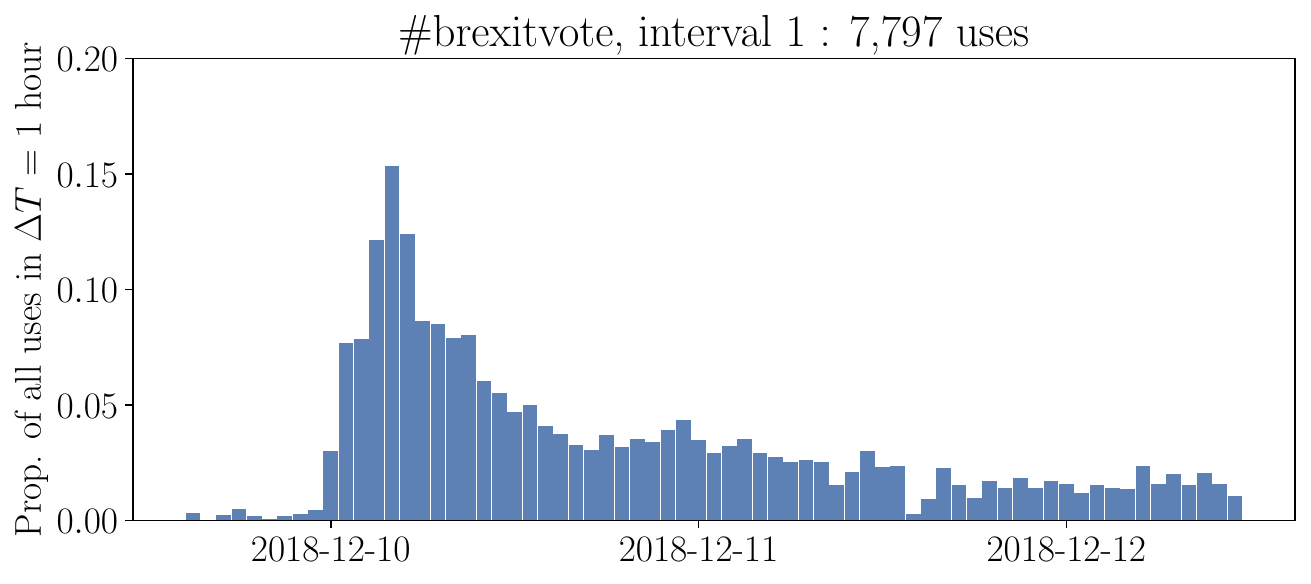}
    \caption{Short-lived peak in hours.}\label{sh:fig:brexitvote_1_hour}
    \end{subfigure} \\
    \begin{subfigure}{0.45\textwidth}
    \includegraphics[width=\linewidth]{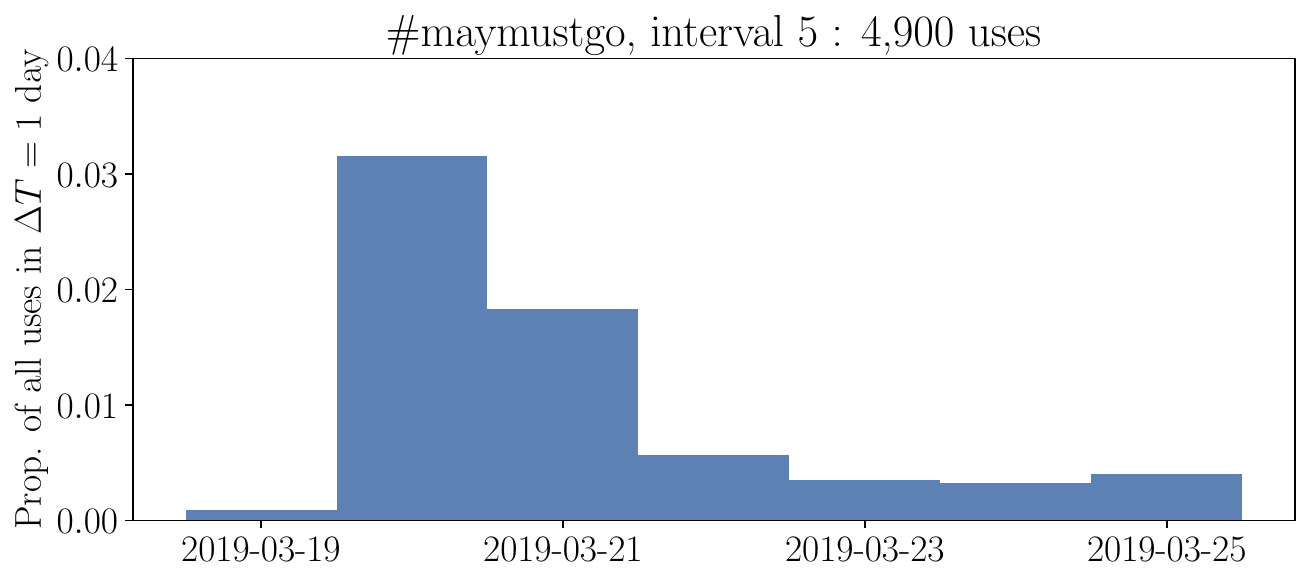}
    \caption{Gradual decline over days.}\label{sh:fig:maymustgo_5_day}
    \end{subfigure} 
    \hfill
    \begin{subfigure}{0.45\textwidth}
    \includegraphics[width=\linewidth]{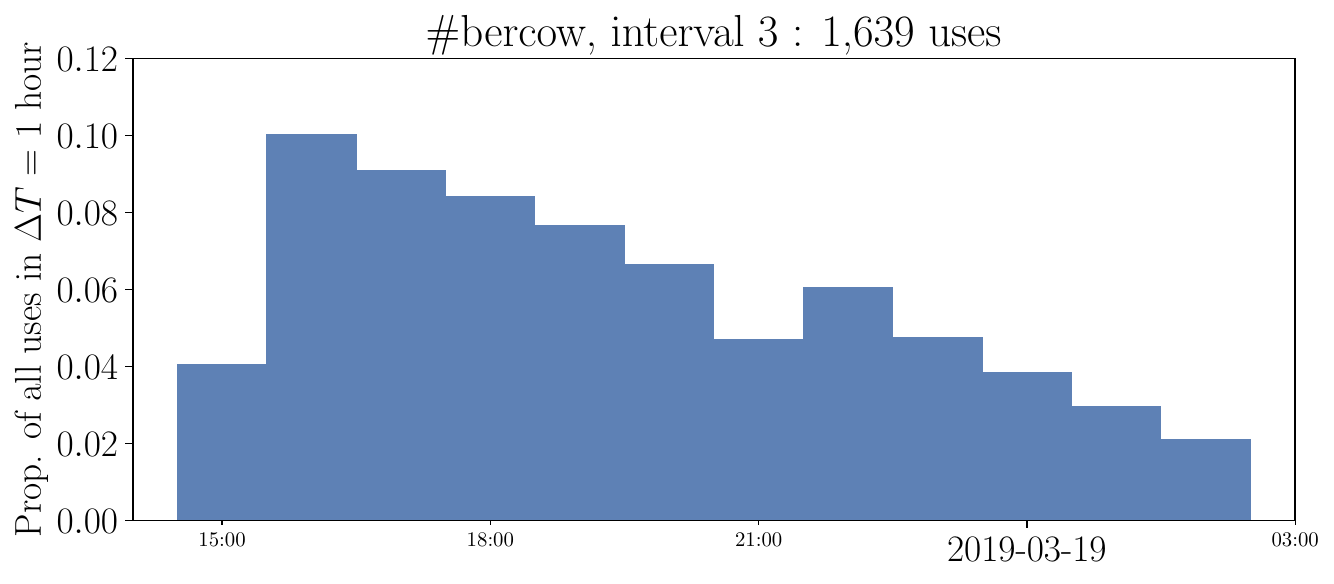}
    \caption{Gradual decline over hours.}\label{sh:fig:bercow_3_hour}
    \end{subfigure}
    \caption{Different hashtags can show similar trends when viewed at different temporal scales (defined by bin width). To compare these directly with each other careful normalisation across volume also needs to be applied. Here it is illustrated how similar behaviour of a short-lived peak and gradual decline can manifest on different time scales.}%
    \label{sh:fig:different_scales}%
\end{figure}

We seek to transform the activity in each event interval such that we can compare events with different time and volume scales. This is an important feature to account for the long-tailed distributions of many metrics related to online social networks~\cite{Barabasi05} which suggests that events with the same underlying generative mechanism can manifest at dramatically different scales. The choice of timeseries resolution can critically alter the visibility of certain trends, and without a consistent timescale for events on social media there remains no obvious choice. Fig.~\ref{sh:fig:different_scales} demonstrates the similarities between events with different scales that can be observed using examples from our dataset. In light of this observation, we define a fair comparison between two intervals as one that reflects common trends of growth and decay, whether a meme spreads through a social group or a large segment of the social network, over a few hours or several days. To achieve this goal, our methodology will not consider time or absolute tweet counts and will instead consider the rate of tweets. Considering tweet rate also helps to resolve independent non-event drivers of tweet volume, such as the day-night cycle, week-weekend cycle, and longer-term overarching trends in platform usage. At a suitable scale, binning resolves some of these factors but the difficulty with determining appropriate scales for multiple events motivates our primary goal. Our approach to manage these factors is detailed in this section.

The first step in our approach assigns each post in the dataset a sequential ID. This enables consideration of rates relative to the collected data which are obscured when using post metadata such as tweet IDs assigned by Twitter (since they are affected by activity globally). Then, given an interval with $n$ posts over an arbitrary temporal width, the scale-independent representation is derived as follows, including an illustrative example using \emph{\#betrayal}.

First, collect the dataset IDs assigned to each post in the interval that uses the topic of interest. 
    \begin{equation}(11207549,11214100,11215455,\dots,12498903,12519628,12523526)\end{equation}
Normalise these numeric IDs such that they span [0,1]. At this point, the value for the $i$th topical post indicates the proportion of all observed posts in the interval required to see $i$ uses of the topic.
    \begin{equation}(0,0.005,0.006,\dots,0.981,0.997,1)\end{equation}
Choose the dimensionality $N$ of the final representation. The example here uses $N=25$. Sample the cumulative distribution function (CDF) of the normalised post IDs at $N+1$ equally-spaced quantiles (including 0 and 1) determined by the chosen dimensionality.
    \begin{equation}(0,0.042,0.069,\dots,0.841,0.939,1)\end{equation}
Calculate the difference between the sampled quantiles, i.e. the proportion of observed posts required to reach the next quantile.
    \begin{equation}(0.042,0.027,0.02,\dots,0.045,0.098,0.061)\end{equation}
Calculate the post rate between quantiles by inverting the proportion of assigned post IDs required to reach the next quantile.
    \begin{equation}(23.7,36.8,48.5,\dots,22.3,10.2,16.4)\end{equation}

We are left with a representation of length $N$ of how the usage rate for the topic changes across the interval. More concretely, given the choice $N=50$, the vector elements in $[0,1]$ relate to the proportion of all posts observed in the time period required to give the next $2\%$ of total topic engagement, i.e. the rate of all posts that include the topic we are investigating. We invert the values in the representation to aid interpretation such that larger values indicate a higher frequency of posts containing the topic of interest.

A point of clarification should be made about the distinction between the number of times a topical indicator occurs and the number of posts a topic appears in. The technique we propose here is compatible with either approach depending on the context in question. When looking at tweets, the total length was constrained to 280 characters which limits the capacity for repetition of a hashtag. In the case of longer texts, other topical indicators (e.g. news articles and named entities) the usage rate within a typical document should be considered carefully relative to the total number of documents. For the purposes of this section, we allow for multiple instances of a topical indicator within a single post. Given the high volume and short length of the tweets considered, we expect that the difference is only impactful in rare cases exhibiting anomalous behaviour, for example a small number of tweets each using the same hashtag many times.

\begin{figure}[!ht]
\centering
\begin{subfigure}{\textwidth}
\centering
    \includegraphics[width=0.65\linewidth]{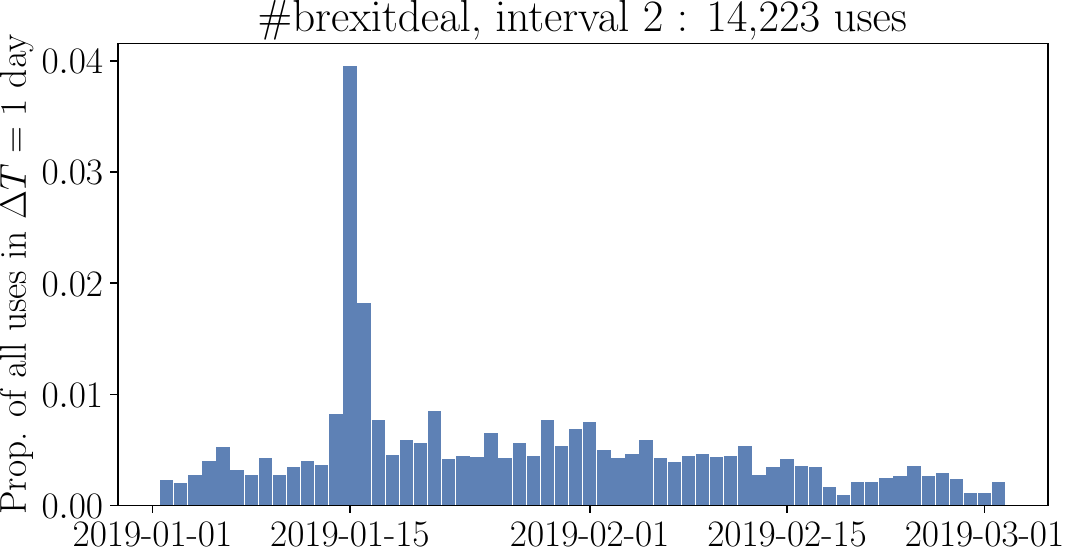}
    \caption{A portion of the hashtag's activity is considered, in this case around an increase in the relative usage rate.}\label{sh:fig:schem_timeseries}
    \end{subfigure}\\
\begin{subfigure}{\textwidth}
\centering
    \includegraphics[width=0.65\linewidth]{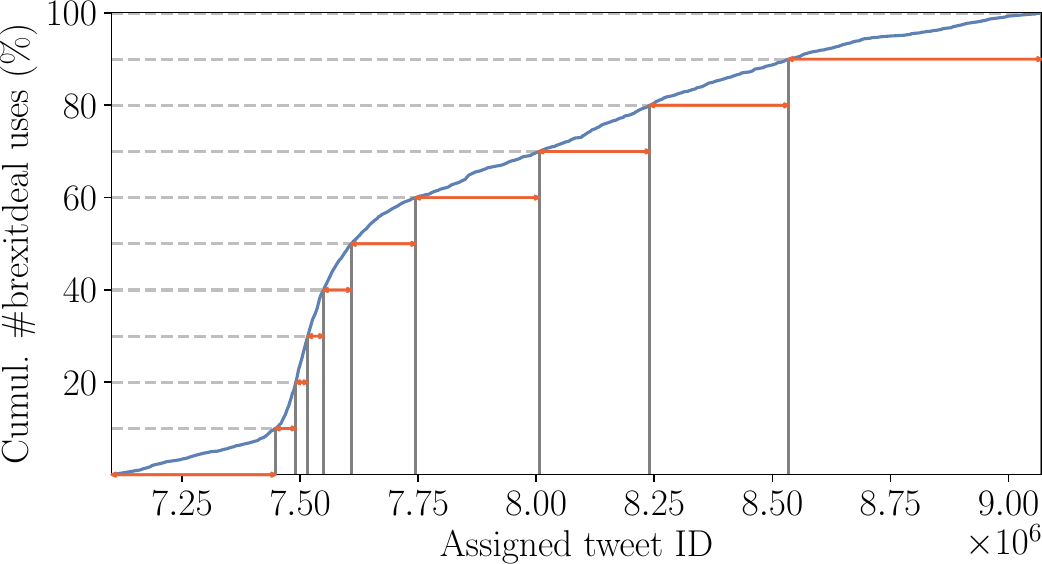}
    \caption{Quantiles in the CDF of dataset tweet IDs are calculated.}\label{sh:fig:schem_cdf_tweet_ids}
    \end{subfigure}\\
\begin{subfigure}{\textwidth}
\centering
    \includegraphics[width=0.65\linewidth]{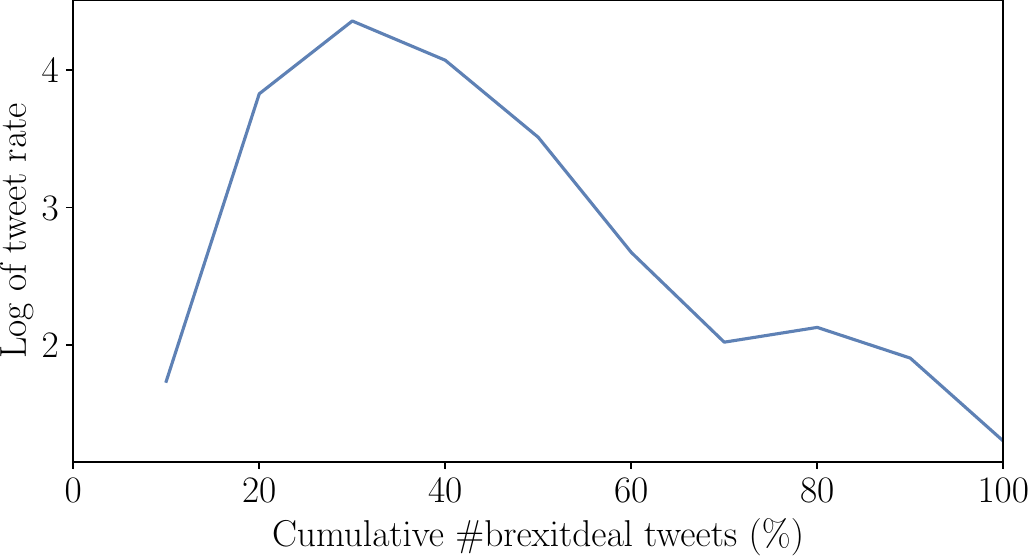}
    \caption{The quantiles are used to measure the activity rate on the hashtag.}\label{sh:fig:schem_vec}
    \end{subfigure}
\caption{In order to resolve the differences in number of tweets between different timeseries segments we apply our scale-independent representation. For reference, the daily binned timeseries is shown in Fig.~\ref{sh:fig:schem_timeseries}. In Fig.~\ref{sh:fig:schem_cdf_tweet_ids} we calculate the CDF and find the tweets observed between successive quantiles of interest (red arrows). We use these tweet IDs to produce a vector of the desired length for each interval as shown in Fig.~\ref{sh:fig:schem_vec}. Here we set $N=10$ for visual clarity, but use $N=50$ for all subsequent analysis.}
\label{sh:fig:cdf_schematic}
\end{figure}

To demonstrate this representation in practice, Fig.~\ref{sh:fig:schem_timeseries} shows the hourly proportion of all tweets using hashtag \emph{\#brexitdeal}. We transform the timeseries into cumulative usage of each hashtag against all tweets in the dataset, shown in Fig.~\ref{sh:fig:schem_cdf_tweet_ids}.
For this example, we choose $N=10$. The dashed horizontal lines and solid vertical lines indicate the quantile values sampled. Finally the sampled IDs are normalised, differenced (red arrows in Fig.~\ref{sh:fig:schem_cdf_tweet_ids}) and inverted to obtain the activity shape shown in Fig.~\ref{sh:fig:schem_vec}.

Through the normalisation procedure and defining the final length of the representation, this methodology achieves our goal of facilitating direct comparison between different scales. The most important of these adjustments is the consideration of relative tweet rates rather than absolute counts since these rates are comparable across periods of different total activity. Additionally, they are unaffected by temporal activity rhythms that can see notable changes in absolute counts under binning methods unless they affect the topic of interest differently (a pattern we wish to recover). One further advantage that this methodology presents is in the level of detail with which different periods of real time are viewed. Sampling from the CDF compresses periods of little activity and extracts more data points from periods of higher activity. This feature enables the representation to focus more on the main event period and is therefore less reliant on careful definition of the interval limits. This attribute is an important differentiating factor between our new approach and existing timeseries methods with a given temporal resolution. Where the binning methods consider a series of time periods equally, our scale-independent representation automatically highlights periods featuring the majority of activity regardless of their time span relative to the interval width.

This method requires the choice of a single parameter, the number of quantiles. We tested $N=25$, $N=50$ or $N=100$ quantiles and found that $N=50$ gave a good balance between smoothing the local noise and still retaining enough detail to characterise activity patterns across multiple types of behaviour. Provided that a sufficient minimum threshold is passed, we expect the number of quantiles to have a limited effect on the overall trends displayed. Intuitively, since we are counting the number of tweets required to see the next quantile of hashtag use, increasing the number of quantiles will reduce the component values in the final representation since we measure more intervals of fewer tweets, but the shape of the activity profile should remain similar.

\subsection{Observing scale-independent representations in social media data}\label{sh:results}

\subsubsection{Data}
In this paper we make use of a dataset collected in real time from the Twitter Streaming API using the list of keywords in Table~\ref{sh:API keywords}, selected to capture a range of opinions and themes about the Brexit process and negotiations between the United Kingdom and the European Union. The collection ran between 1st January 2018 and 26th June 2019 and includes 12,545,942 tweets. This period included a few short outages but since our experiments consider periods of activity within this collection they are not expected to affect our results. In order to extract signals of attention, we rely on the hashtags included in the tweets. In the absence of any specific measures of collective attention, we treat these as intentional indicators of the content and terms that users are not only aware of, but also sufficiently invested in to tag in their own posts. Over the 18 month period we collect 950,508 unique hashtags from 1,333,403 users. 

There are important considerations when constructing a dataset for the analysis we propose. Since the scale-independent representation investigates the rate of topical posts relative to the ongoing rates of all posts, the dataset should not be too tightly defined by a single topic used for subsequent analysis. Our choice of topic in the Brexit debate presents a wide range of related themes to avoid this issue of hyper-focus. The use of keyword terms however means that any exploration of their patterns should be aware that there may be inherent skew in any trends and introduce an element of selection bias. Our dataset was collected using a wide range of keywords to mitigate this risk, we use hashtags rather than keywords in the tweet text as our topic indicators and primarily focus on terms that were not selected a priori. The only keyword in Table~\ref{sh:API keywords} appearing in our sample of hashtags with sufficient activity is \emph{\#brexit}. While this is a general term in the conversation, its usage as a hashtag varies across the dataset to the extent that patterns are still visible in the scale-independent representation.

\begin{table}
    \centering
    \begin{tabular}{|p{0.18\columnwidth}|p{0.2\columnwidth}|p{0.18\columnwidth}|p{0.22\columnwidth}|}
    \hline
        brexit &  leaveeu & no2eu & voteleave  \\
        euref & betteroffout & voteno & ukineu  \\
        remainineu & stayineu & yes2eu & incampaign \\
        referendum & european union & strongerin & eureferendum \\ \hline
    \end{tabular}
    \caption{The keywords supplied to the Twitter Streaming API to gather the dataset, returning any use in the tweet text (including hashtags).}
    \label{sh:API keywords}
\end{table}

\subsubsection{Peak identification and dividing the timeseries}\label{sh:method:splitting}

For our analysis, we select a sample of 100 hashtags from the 4,278 hashtags used at least 500 times across the 18-month study period. This threshold is low, averaging approximately one tweet per day, but necessary to ensure a minimum level of activity for an event. To consider different events for the same hashtag it was necessary to manually divide the activity for these hashtags into a series of intervals containing either an event or a period of persistent activity. We explored ways to do this algorithmically, but find there are issues with these methods. The natural choice for this activity would be peak detection methods~\cite{Lehmann12, Olteanu15} (see~\cite{Healy15} for further comparisons) or tipping point analysis~\cite{Boulton19}, but there is no guarantee for such methods to identify the complete lifetime of the event (that is, not just the event peak but also its onset and dissipation). We also note that there is no best choice for peak detection algorithms and comparisons of methods have found large variation in accuracy metrics across different topics~\cite{Healy15}. Given these limitations, the sampled 100 hashtags were chosen such that event periods of twice the surrounding baseline rate were clearly identifiable in the day timeseries of tweet counts through visual inspection. Although this process is not feasible for analysis of large datasets, and biases are introduced by choice of bin, this approach is necessary to demonstrate the efficacy of our proposed method, and demonstrate resolution of events over different timescales (as evidenced in Fig.~\ref{sh:fig:different_scales}). Fig.~\ref{sh:brexit splits} illustrates an example timeseries divided into single-event intervals. To ensure a minimum level of ``attention'' in each interval and limit noise, we only consider those containing at least 100 tweets. This leaves 517 intervals for subsequent analysis, which can represent events or the baseline behaviour between them.

\subsection{The importance of time resolution}

As a motivating example, we first highlight how the appearance of the different classes proposed by Lehmann et al.~\cite{Lehmann12} can be heavily influenced by the choice of bin size in the timeseries. In this example, we use the methodology described by Lehmann et al. to identify the single largest peak usage of a hashtag. We then visualise it with hourly, six hourly and daily bins alongside the scale-independent representation we propose.

\begin{figure}
\centering
\begin{subfigure}{0.45\textwidth}
    \includegraphics[width=\linewidth]{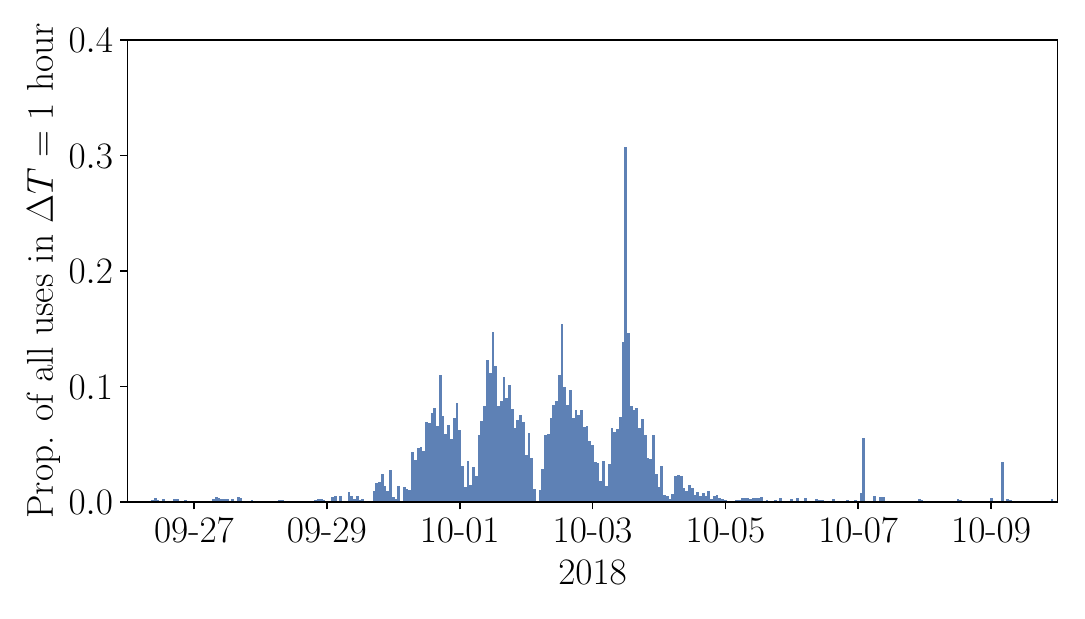}
    \caption{Timeseries with one-hour bin width.}\label{sh:fig:cpc18_a}
    \end{subfigure} 
    \hfill
    \begin{subfigure}{0.45\textwidth}
    \includegraphics[width=\linewidth]{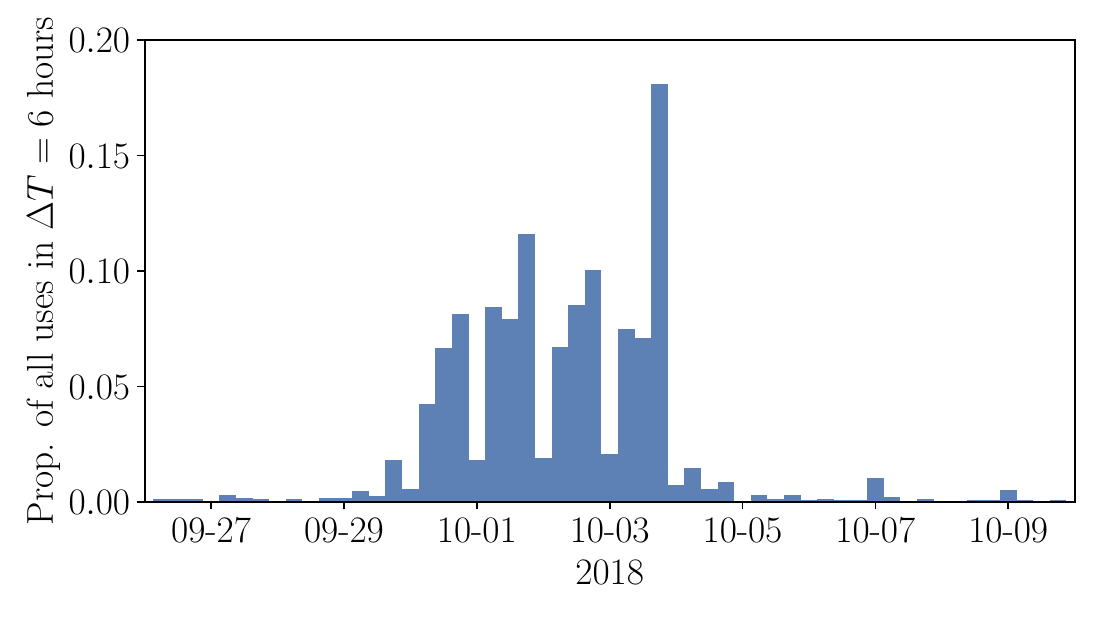}
    \caption{Timeseries with six-hour bin width.}\label{sh:fig:cpc18_b}
    \end{subfigure} \\
    \begin{subfigure}[t]{0.45\textwidth}
    \includegraphics[width=\linewidth]{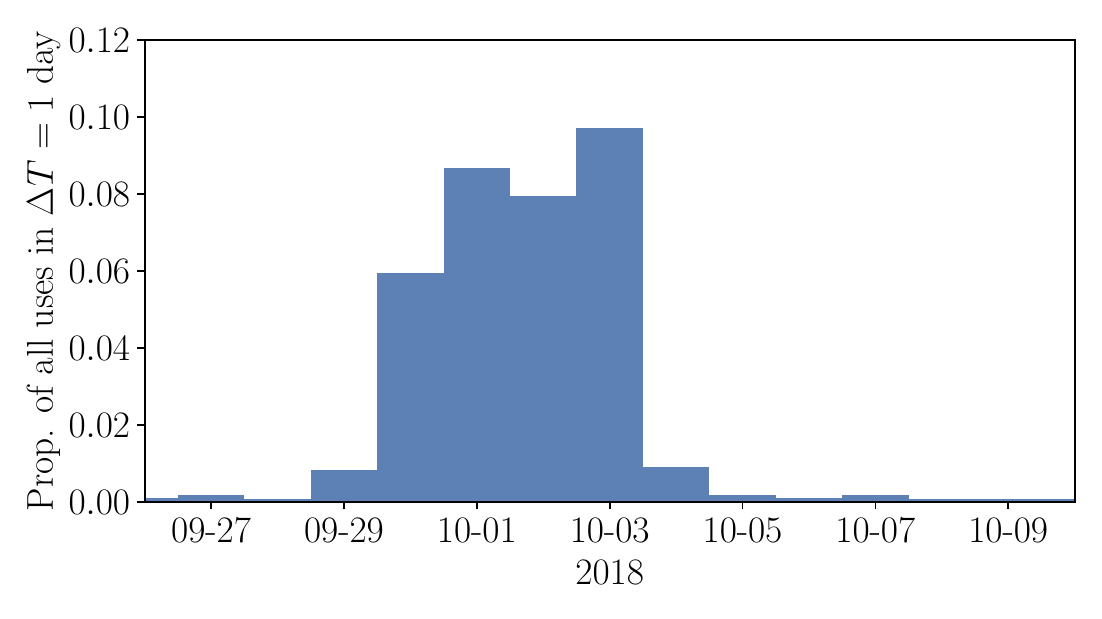}
    \caption{Timeseries with one-day bin width.}\label{sh:fig:cpc18_c}
    \end{subfigure}
    \hfill
    \begin{subfigure}[t]{0.45\textwidth}
    \includegraphics[width=\linewidth]{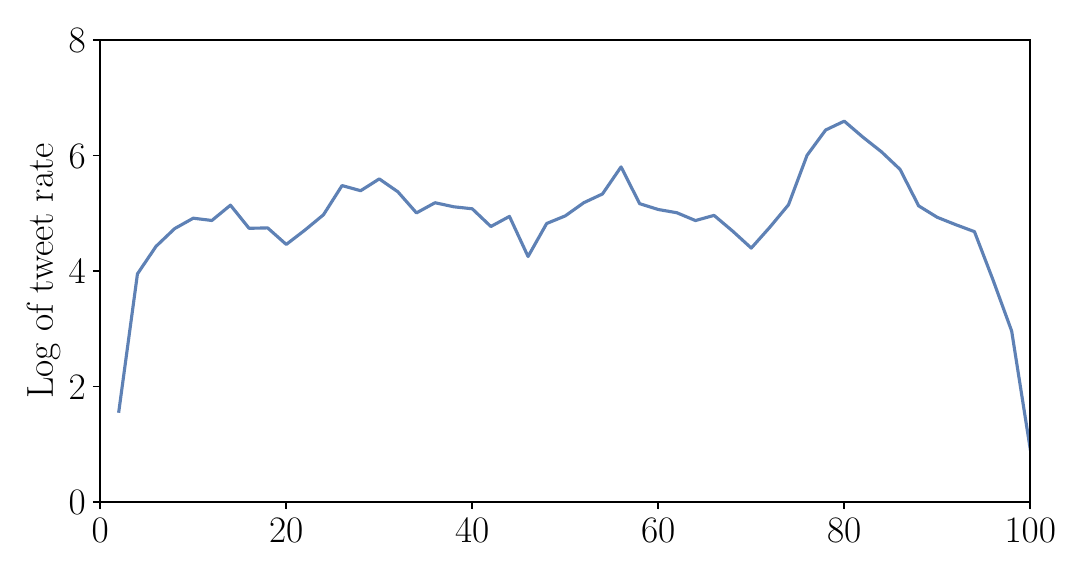}
    \caption{Scale-independent representation of the hashtag activity with $N=50$.}\label{sh:fig:cpc18_d}
    \end{subfigure}
\caption{\emph{\#cpc18}, used to signify attendance or interest in the 2018 UK Conservative party conference. Choosing a bin width of one day presents this period as a single event, gradually building up to a peak. At shorter time resolutions, we see that activity rises and falls across each day of the conference. Fig.~\ref{sh:fig:cpc18_d} shows the scale-independent representation of this period.}
\label{sh:fig:cpc18}
\end{figure}

In Fig.~\ref{sh:fig:cpc18} we explore \emph{\#cpc18}, a hashtag used at this time to show participation or interest in the 2018 UK Conservative party conference. The one-day width bins in Fig.~\ref{sh:fig:cpc18_c} suggest that the period captured as an event by Lehmann et al.'s method is a single event and falls within their ``before and during the peak'' class. At the shorter timescales (Figs.~\ref{sh:fig:cpc18_a},~\ref{sh:fig:cpc18_b}) it is revealed that the period actually covers a series of daily events peaking around the middle of the day as the events of the morning are discussed and anticipating the afternoon's news. The scale-independent representation over this period reflects the daily rhythm of smaller events and furthermore reflects the greater rate of activity seen in the final day of the conference. It should be noted however that the relative sizes of these events are linked to their visibility at different resolutions of the scale-independent representation. In the case of \emph{\#cpc18}, each of the daily events has broadly similar total activity levels, with the final day seeing the most coverage. This trend is reflected in the scale-independent representation, which sees the largest tweet rate in the final day of the period. A similar example for Lehmann et al.'s ``peak only'' class is shown in Fig.~\ref{sh:fig:esthermcvey}.

\subsection{Characteristic shapes}\label{sec:shapes}

We continue by examining the different profiles found under periods of increased hashtag usage rate to understand how the four classes of Lehmann et al. (``before and during the peak'', ``during and after the peak'', ``symmetrically around the peak'' and ``peak only''~\cite{Lehmann12}) translate to our new methodology. We find four broad shapes of profile, three of which match closely to similar classes proposed by Lehmann et al.

\begin{figure}[!ht]
    \centering
    \begin{subfigure}{0.45\textwidth}
    \includegraphics[width=\linewidth]{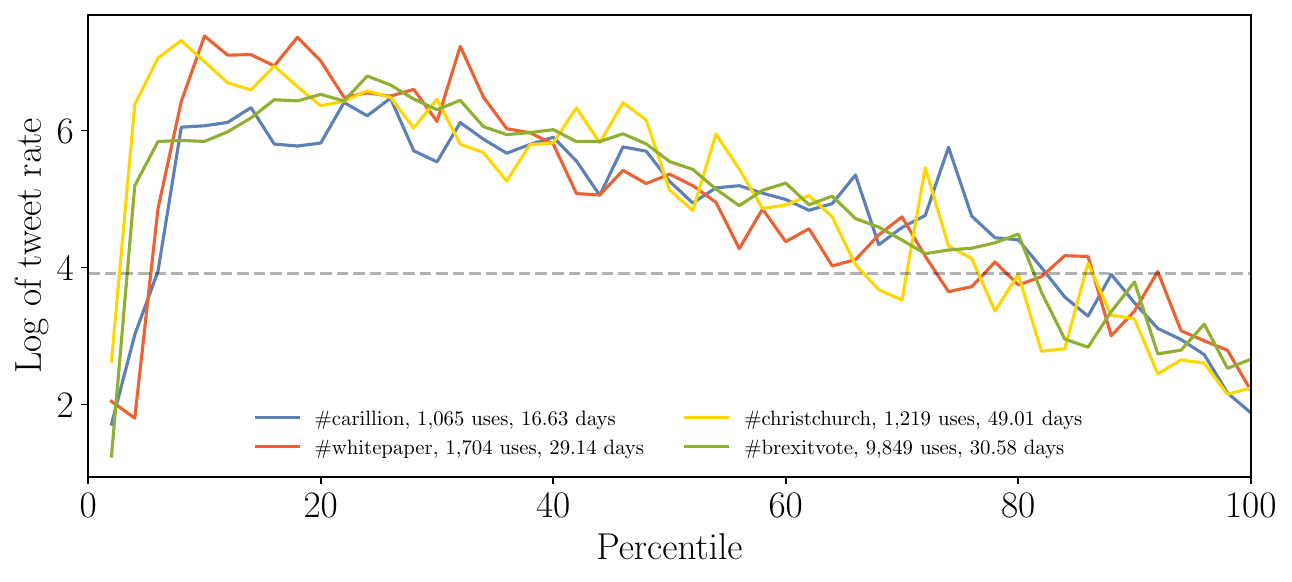}
    \caption{Right-tailed}\label{sh:fig:right-tailed}
    \end{subfigure} 
    \hfill
    \begin{subfigure}{0.45\textwidth}
    \includegraphics[width=\linewidth]{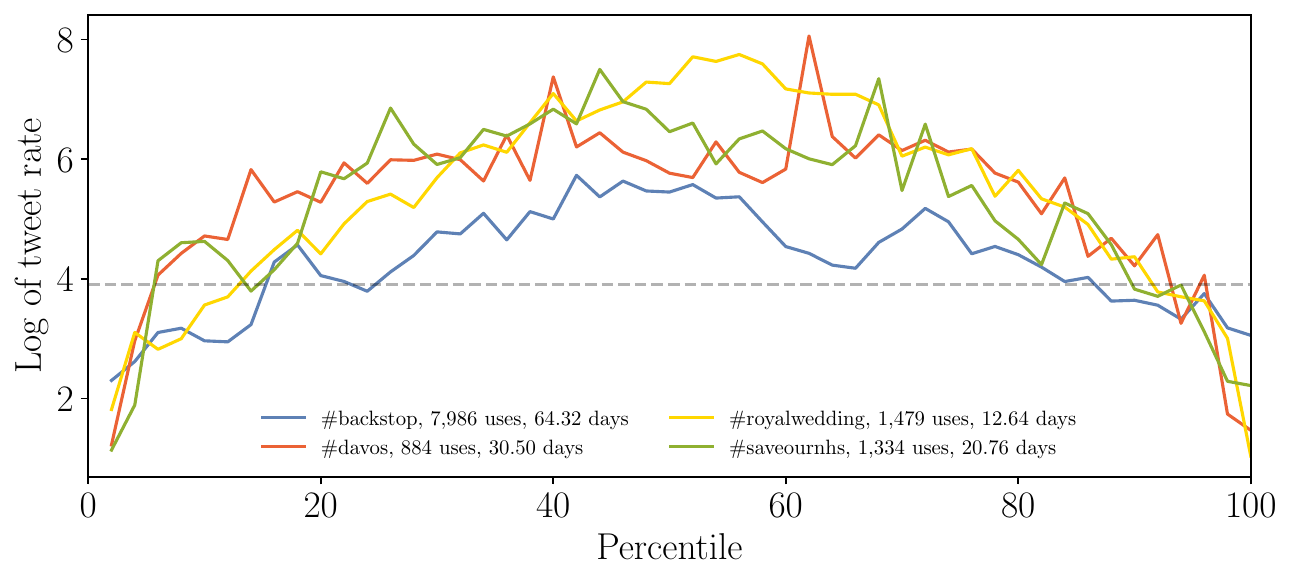}
    \caption{Arch-shaped}\label{sh:fig:arch-shaped}
    \end{subfigure} \\
    \begin{subfigure}{0.45\textwidth}
    \includegraphics[width=\linewidth]{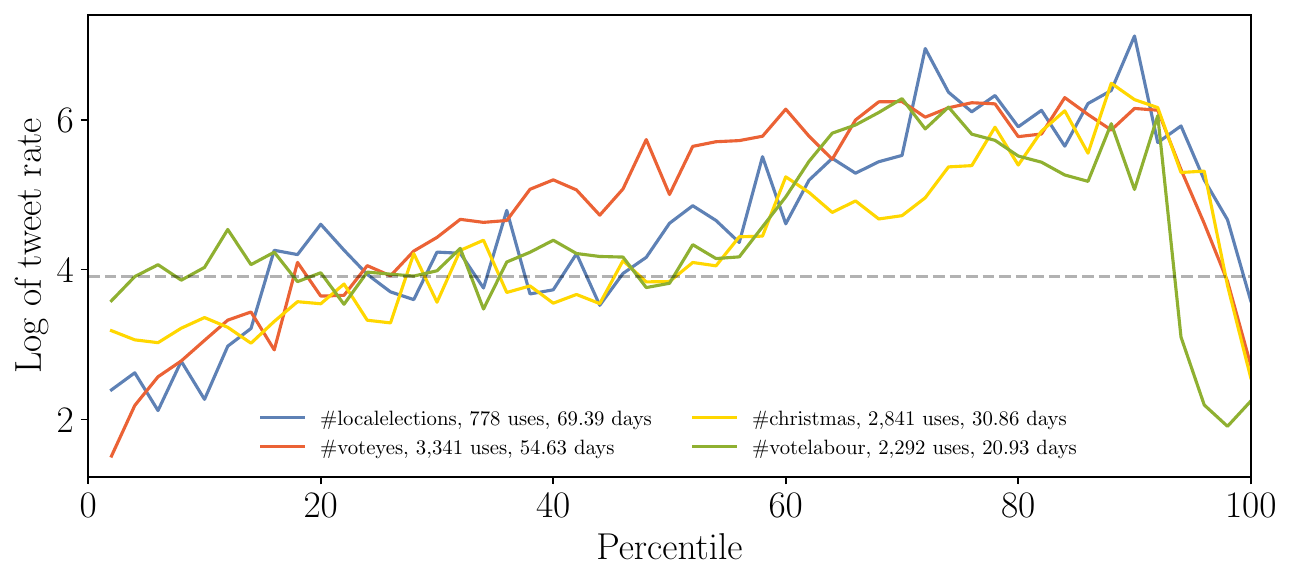}
    \caption{Left-tailed}\label{sh:fig:left-tailed}
    \end{subfigure}
    \hfill
    \begin{subfigure}{0.45\textwidth}
    \includegraphics[width=\linewidth]{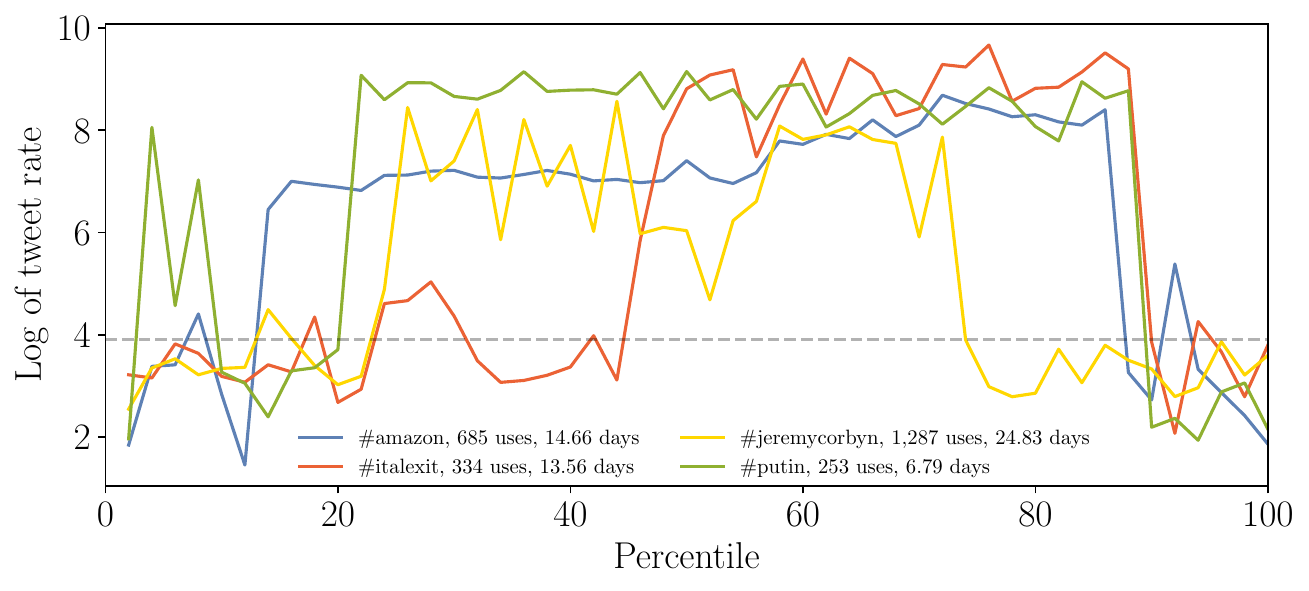}
    \caption{Abrupt shift}\label{sh:fig:abrupt shift}
    \end{subfigure}
    \caption{The four characteristic shapes found around increased hashtag usage rate. Different coloured profiles indicate different events. The dashed line approximates the value of constant activity evenly distributed through the lifetime. a) The right-tailed shape indicates events with a sudden increase in attention followed by a gradual decrease. b) The arch-shaped profile indicates events with both a gradual increase and a gradual decrease in attention. c) The left-tailed shape indicates events with a gradual increase in attention before a sudden decrease. d) The abrupt shift shape indicates a sudden transition in and out of a period of steady, heightened attention.}%
    \label{sh:fig:representative shapes}%
\end{figure}

In Fig.~\ref{sh:fig:right-tailed} we present examples of the first shape identified in periods of increased activity. This \emph{right-tailed} profile is typical of unanticipated events such as breaking news stories or unforeseen natural disasters. The scale-independent representation increases abruptly and quickly peaks before gradually decreasing and returning to a background rate. Here we recover the ``during and after the peak'' class from Lehmann et al.

In Fig.~\ref{sh:fig:arch-shaped} we present examples of an \emph{arch-shaped} profile characterised by a gradual increase of hashtag usage rate to the peak followed by an, often roughly symmetrical, decline to the background use rate. This profile shape corresponds to Lehmann et al.'s class ``symmetrically around the peak'' and constitutes many anticipated events which allow for both steady growth of attention, and sustained interest after the peak of activity, particularly scheduled announcements, interviews or holidays.

In Fig.~\ref{sh:fig:left-tailed} we provide examples of a \emph{left-tailed} profile characterised by a gradual increase in hashtag usage rate to a peak, beyond which interest is not maintained, and usage rapidly returns to the background rate. This shape is the mirror complement of the right-tailed shape, common with events that have building anticipation beforehand but retain little relevance, or are quickly relabelled after the event, such as an anticipated interview which is quickly usurped by discussion of the talking points. This shape most closely relates to the ``before and during the peak'' class observed by Lehmann et al.

The final characteristic shape we observe is shown in Fig.~\ref{sh:fig:abrupt shift}. Here we find periods of greatly increased hashtag usage rates with sudden changes between a normal state and an event state. We call this an \emph{abrupt shift} profile. This shape is distinct from the three other characteristic shapes shown in Fig.~\ref{sh:fig:representative shapes} by having no gradual change in rate on either side of the peak. We observed this behaviour in hashtags such as \emph{amazon}, \emph{italexit}, \emph{putin} and \emph{trump}. It is difficult to intuit a type of organic collective interaction that could lead to this hashtag usage rate profile. Inspection of usage of the hashtags which create rate profiles of this type suggests commercial interests or spamming behaviour, heavily implying automated ``bot'' contributions to the dataset. Some examples of such tweets are shown in Table~\ref{sh:tab:shift examples}. Lehmann et al. do not identify this as a distinct phenomenon, although in the case where the shift occurs within a single day, it would be miscategorised in their ``peak only'' class, that is an event with both an abrupt increase and decline around a focused peak. 

More generally, we find no clear indication of the ``peak only'' class being recovered under our representation. Certainly our dataset contains events in the above categories which fall within a 24-hour span, but our non-parametric approach has shown in many cases that the underlying event behaviour falls within the other three categories. This is not surprising since classification of events as ``peak only'' is highly dependent on the choice of bin width and we have seen that collective attention events lack a characteristic timescale, see e.g. Fig.~\ref{sh:fig:different_scales}.

\subsection{Discrete categories or a continuum?}

\begin{figure}[!ht]
    \centering
    \includegraphics[width=0.7\textwidth]{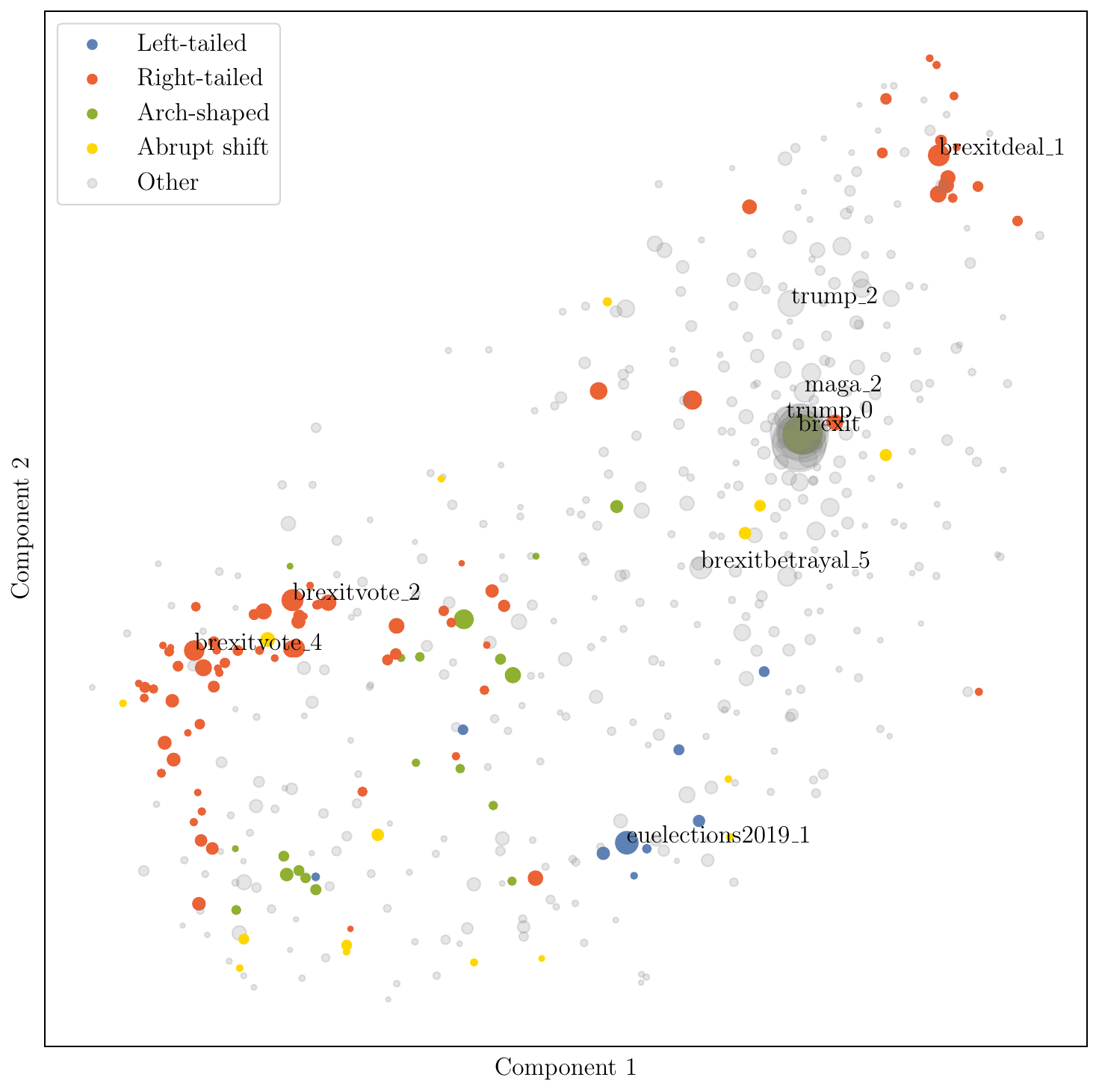}
    \caption{Projection of the scale-independent representations of hashtag intervals using the two dominant components under t-SNE analysis. Point colours denote the type classification by the authors and point size is proportional to the square root of total usage in the interval. Grey, low opacity points indicate periods of activity which did not clearly show any of the four characteristic shapes discussed in Section~\ref{sec:shapes}. Hashtag and interval number labels are provided for intervals with more than 22,500 tweets, and the label for \emph{\#brexit} applies to all intervals which overlap in the t-SNE projection. This figure excludes hashtags with anomalous behaviour from repeated usage in a single tweet, whose representations were not compatible with t-SNE projection. We see that the left-tailed, arch-shaped and right-tailed profiles approximate frontiers in this projection and suggest a transition of behaviours between activity shapes.}
    \label{sh:TSNE}
\end{figure}

The examples of the shapes presented previously are clear cut, but many of the hashtags studied suggest that the scale-independent representation produces a continuum of shapes rather than discrete classes as observed by Lehmann et al. Fig.~\ref{sh:TSNE} projects the scale-independent representations for different intervals into the two principal t-distributed stochastic neighbour embedding (t-SNE) components~\cite{vanderMaaten08}. t-SNE is a conditional probability-based method for embedding higher dimensional points in lower dimensional space with additional optimisation for use on large datasets. The embedding aims to arrange points such that it maximises the likelihood of point $x_i$ choosing other points as neighbours given a Student's $t$-distribution centred on $x_i$. There is a separation of left-tailed and right-tailed events, but rather than a distinct boundary, there is a transition through arch-shaped profiles in projected space. We argue that this points to approximate frontiers in the projected space between these three types of behaviour. Under this projection the abrupt shift class is not clearly distinguished from the three other classes. The activity periods that do not clearly display one of the characteristic behaviours are well-distributed throughout this projection and further support the argument for a continuum of behaviours. Four hashtag intervals in this group are not shown in Fig.~\ref{sh:TSNE} due to discontinuities in their representation. This occurred when single tweets included many uses of the same topic and total activity is otherwise low. Since these examples emerge only in the extremes of the activity distributions they represent atypical behaviour on Twitter. Fig.~\ref{sh:TSNE} reveals unusual behaviour in \emph{\#brexit} compared to the other topics considered. The activity patterns in each of the studied intervals are close in projected space which suggests that many similar events occur in this topic. Several other hashtags exhibit this behaviour among certain periods (notably \emph{\#brexitvote} and \emph{\#trump}) but lack the consistency over time that we see in \emph{\#brexit}.

\begin{figure}%
    \centering
    \begin{subfigure}{0.45\textwidth}
    \includegraphics[width=\linewidth]{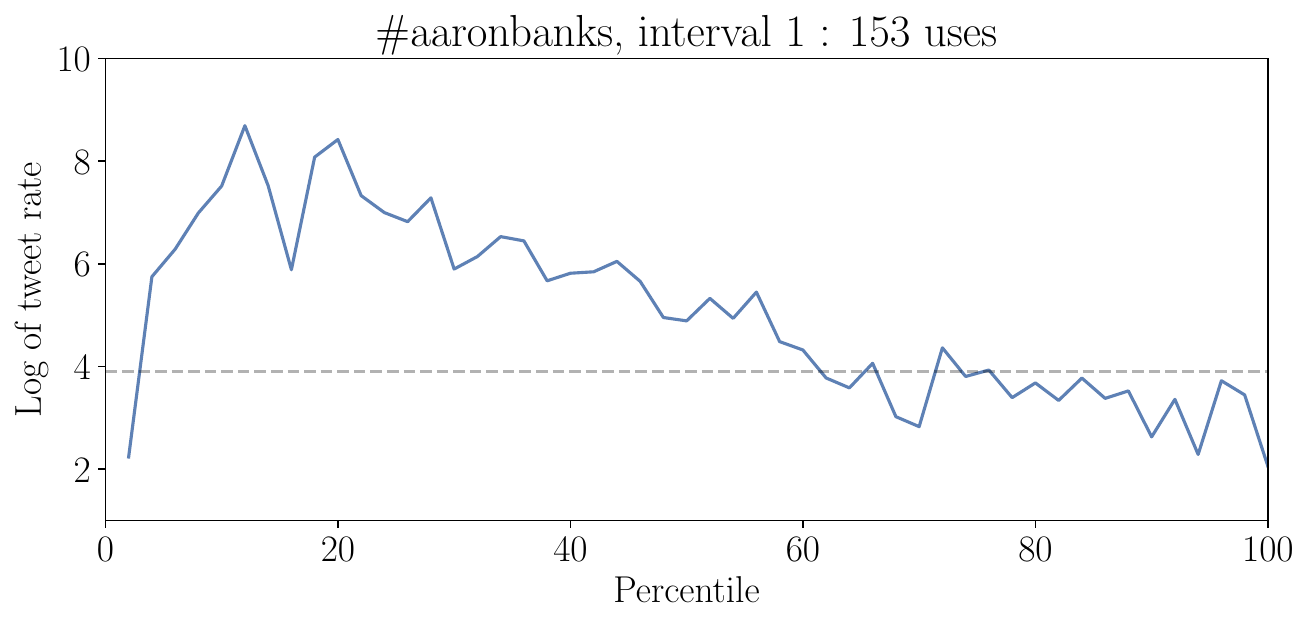}
    \caption{Short-lived peak in days.}\label{sh:fig:aaronbanks_1_vec}
    \end{subfigure}
    \hfill
    \begin{subfigure}{0.45\textwidth}
    \includegraphics[width=\linewidth]{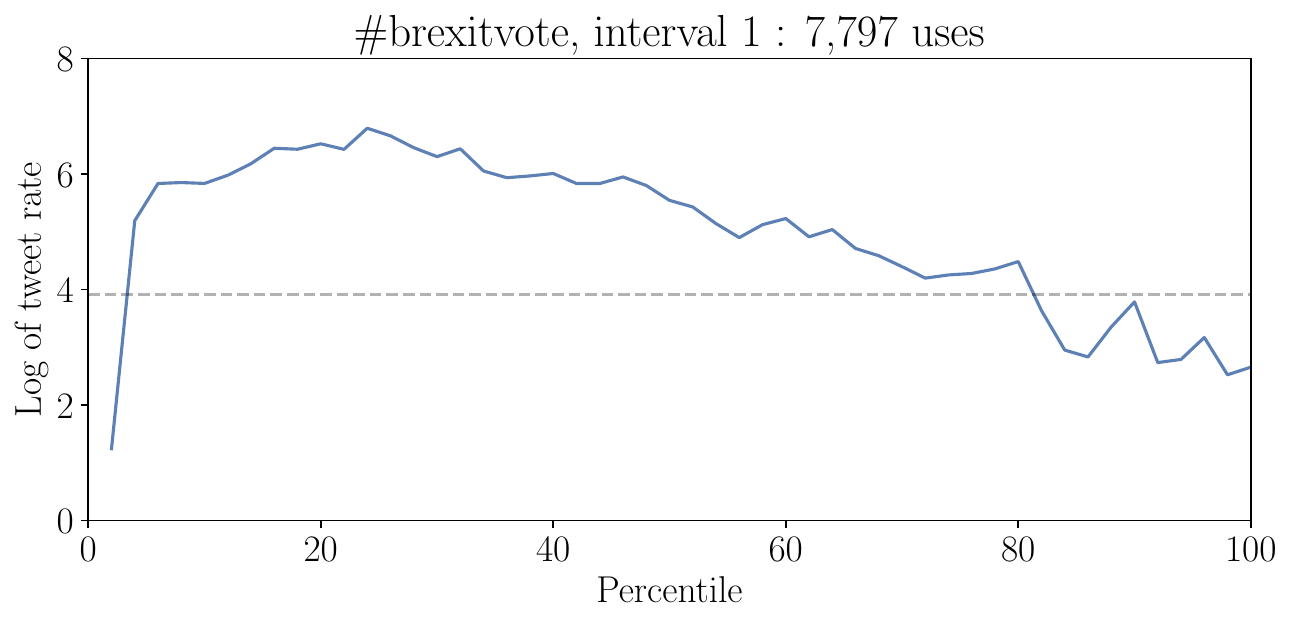}
    \caption{Short-lived peak in hours.}\label{sh:fig:brexitvote_1_vec}
    \end{subfigure}\\
    \begin{subfigure}{0.45\textwidth}
    \includegraphics[width=\linewidth]{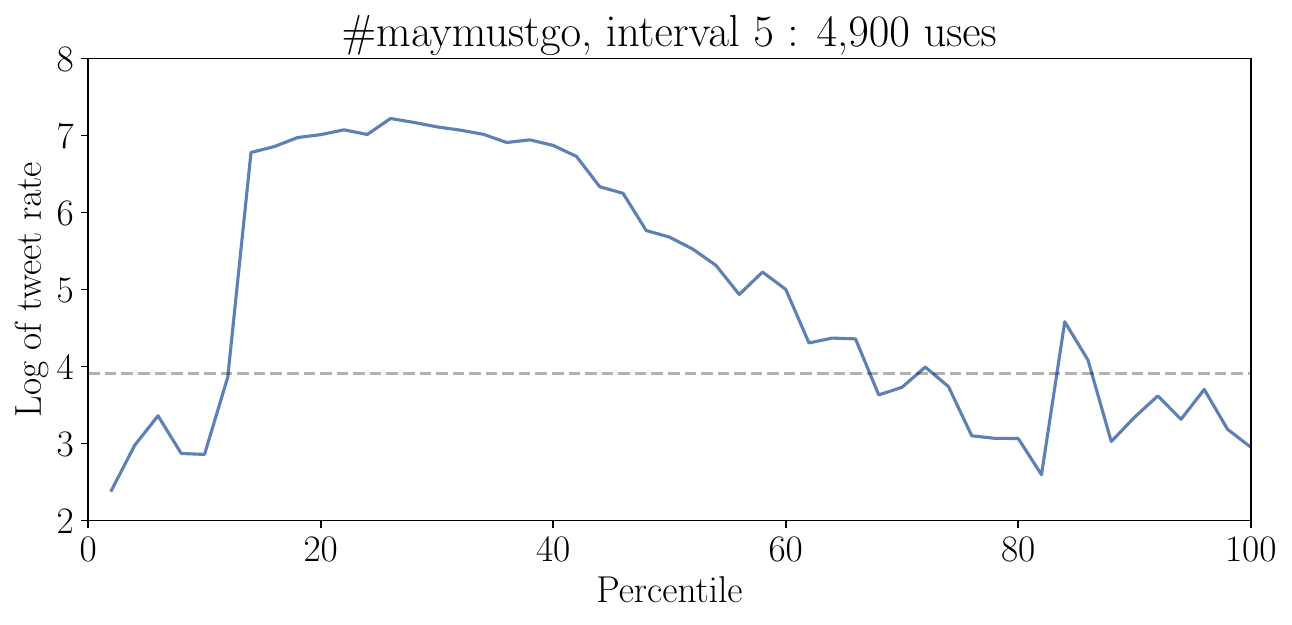}
    \caption{Gradual decline over days.}\label{sh:fig:maymustgo_5_vec}
    \end{subfigure} 
    \hfill
    \begin{subfigure}{0.45\textwidth}
    \includegraphics[width=\linewidth]{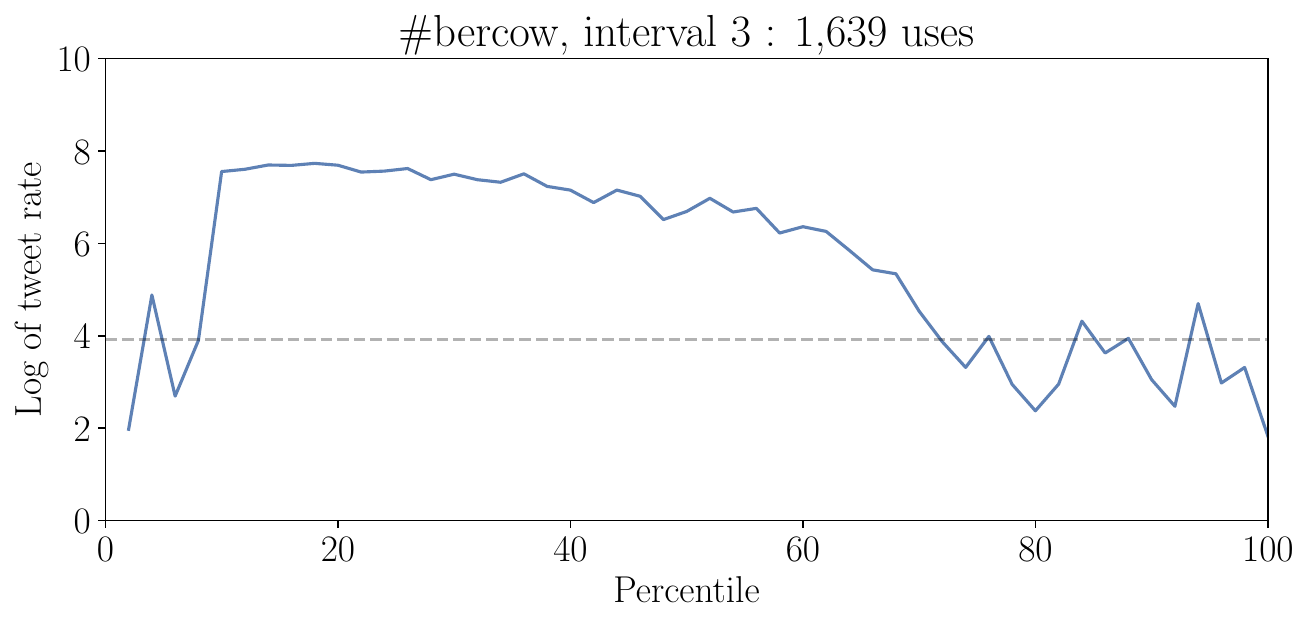}
    \caption{Gradual decline over hours.}\label{sh:fig:bercow_3_vec}
    \end{subfigure}
    \caption{The scale-independent representations for the four examples from Fig.~\ref{sh:fig:different_scales}.}%
    \label{sh:fig:different_scales_vecs}%
\end{figure}

In Fig.~\ref{sh:fig:different_scales_vecs} we calculate the scale-independent representations of the four example segments from Fig.~\ref{sh:fig:different_scales}. We can see that Figs.~\ref{sh:fig:aaronbanks_1_vec},~\ref{sh:fig:brexitvote_1_vec},~\ref{sh:fig:maymustgo_5_vec} and~\ref{sh:fig:bercow_3_vec} all show the same right-tailed profile shape despite their differences in the binned timeseries. This example demonstrates the robustness of the scale-independent representation to variations in the intervals studied. Each of the four examples here can be broadly characterised as right-tailed, but there are subtle differences between the shapes of the short-lived peaks and the gradual declines. 

Of the periods identified in the hashtag usage, a total of 517 intervals of length at least 100 tweets were found. 78 were identified as right-tailed, 20 as arch-shaped, 10 as left-tailed, 16 as abrupt shifts, 32 as uniform (i.e. close to the level of constant use) and 361 did not clearly fall into one of the other categorisations. Identifying several intervals as uniform activity adds another useful attribute to the method. In these cases, the interval represents a period of steady activity between two periods of change. With the ability to distinguish these periods, it is not necessary to identify event intervals during the splitting step, only the boundaries at which the activity profile changes. The fact that the majority of events do not clearly fall into one of the categories under visual inspection adds further support to the notion that these characteristic shapes are points on a continuum rather than discrete classes. Most of the intervals that were difficult to classify covered periods where the daily rate of hashtag usage oscillated rapidly, fell between two of the classes or contain multiple events that are clarified by the scale-independent representation. While this last situation arises from the limitations of the manual splitting process, it does demonstrate the potential for our approach to distinguish multiple events within a single interval.

\section{Mechanistic properties of collective attention}\label{sec:model}

Through the investigation of how attention to topics in the Brexit debate on Twitter evolve over time, we have seen that the scale-independent representation we propose presents a new perspective on how collective attention emerges in scale-free social systems. In the previous section we have seen that the proposed representation recovers behavioural patterns that are both similar and dissimilar to those found in previous works and illustrated that there are similarities in patterns across time and volume scales. We have also seen that the discrete categories of activity found previously may instead represent positions on a continuum which are obfuscated in many cases by multiple related events within a given interval producing multiple overlapping activity patterns. Moreover the independence of these findings to time and volume scales suggests that the patterns we reveal may instead reflect universal mechanisms that are common to processes of attention in all social systems. In this section we will build a simple agent-based model to represent this behaviour and investigate the factors that determine which shape of representation we observe.

\subsection{Defining a simple agent-based model for collective attention}

Our modelling framework is designed to reflect a wide range of social systems in the modern world, where individuals have awareness of issues beyond their local connections. This concept is designed to represent social media platforms (which commonly have feeds that highlight popular content in addition to that served from selected contacts), but will remain sufficiently general to represent a diverse set of circumstances (e.g. a geographically-defined social network with broader issue awareness through news broadcasts). In addition to this split between local and global issue awareness, we simplify the topic space to consider only the distinction between on- and off-topic messages.

Previous attempts in this direction have included similar limitations to Lehmann et al.'s original work~\cite{Lehmann12}. The closest effort to our intended goal, produced by Huynh et al.~\cite{Huynh15}, proposed an agent-based model for engagement patterns with a given topic. This model again focused on post frequency within a given time interval to produce the same timeseries behaviour seen by Lehmann et al. and explain this behaviour using two modelling parameters. In addition to the same potential bias introduced by the choice of a particular timescale (and frequency at that timescale), this modelling approach does not consider the usage of other topics beyond the one of interest. This limitation in particular renders the model unsuitable for the application of our new scale-independent representation (which requires off-topic activity) and motivates our decision to design a new model.

Here we outline an agent-based model simulating the adoption and diffusion of a given idea or topic through a social network. Inspired by the patterns seen in the Twitter data discussed in the previous section we focus on the effects of two factors influencing the activity produced by the model: the functional form determining wider attention to the topic of interest over time and the presence of manipulative users who repeatedly post the same content for a given period of activity.

At each time step agents report one of three states: 0, indicating that they are not active at this time; 1, indicating that they are active but posting off-topic content; and 2, indicating that they are active and posting on-topic content. Note that this division into on- and off-topic content means that this model does not consider competing topics individually but instead treats all such messages as identical and ignores any variations within the range of possible topics covered. In addition, this consideration of on- and off-topic states means that in applying the scale-independent representation we must allow only a single topic for each post in contrast to the Twitter analysis in the previous section.

Our modelling framework consists of two agent types, representing authentic and spamming users, which use different mechanisms for choosing their state at each time step. The spamming agents (detailed in Algorithm~\ref{alg:botnet_pseudocode}) are controlled by parameters indicating a start and stop time for coordinated activity. During this period of coordinated activity, our model assumes that all spamming agents will post on-topic content exclusively, and outside of this period they will not post.

\begin{figure}
    \centering
    \includegraphics[width=\linewidth]{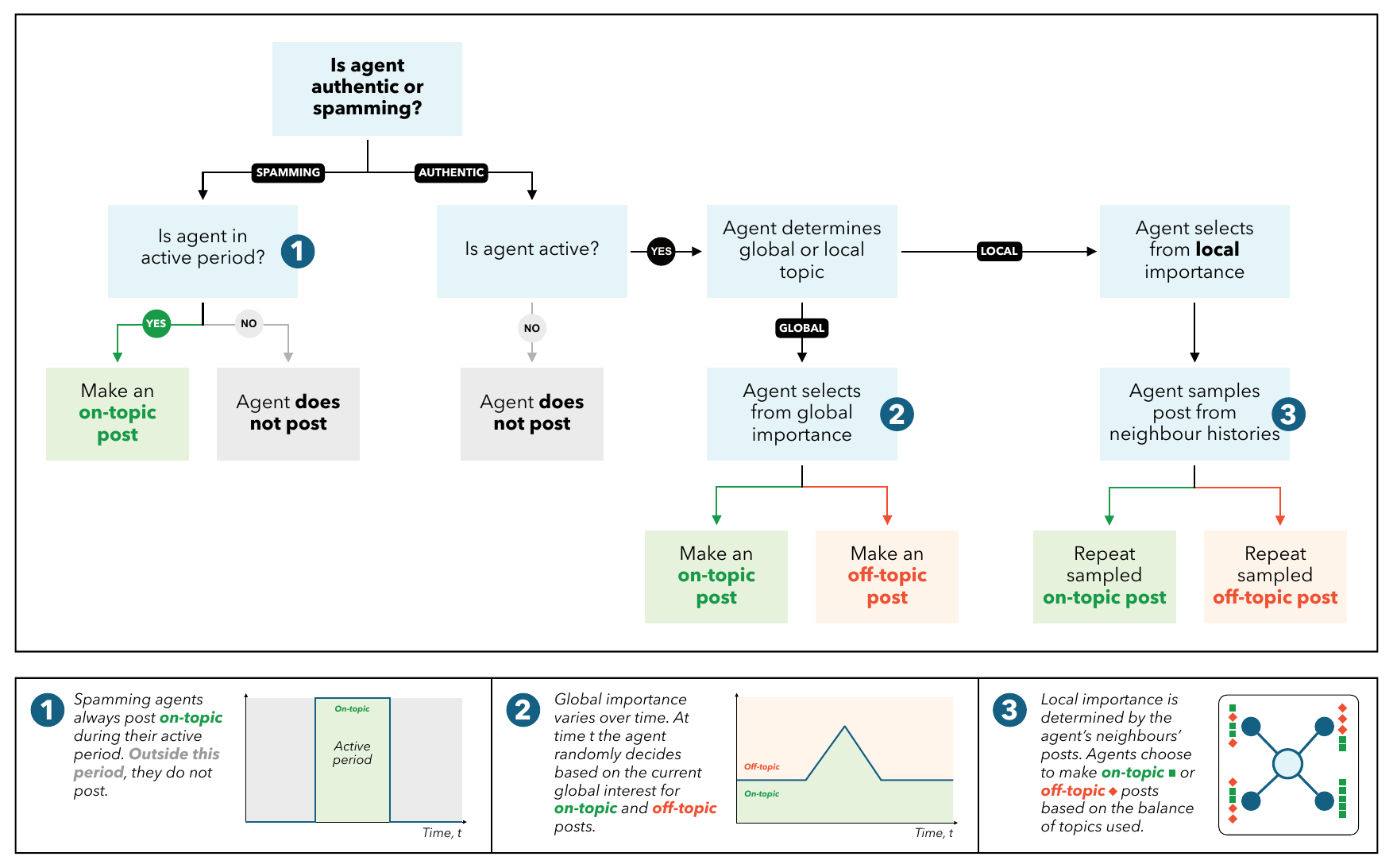}
    \caption{Schematic overview of agent types in the agent-based model. The model includes two types of agent, authentic or spamming who exhibit different behaviours. Spamming agents repeatedly post on-topic during their active period (1). Authentic agents are designed to mimic the behaviour of human users and select their topic from a global trending feed determined by global importance (2) or from those posted by their social network (3).}
    \label{fig:model_schematic}
\end{figure}

The authentic agents in our model follow expected patterns of human user behaviour, choosing what to post based on the topics used by their social network and those that are important globally. These agents are defined by a series of preferences: how active they are on average; their preference for locally rather than globally important topics; and the global visibility of the topic of interest. We capture the external dynamics of global importance of the topic of interest in a time-dependent function to allow for a wide range of exogenous behaviours to be reflected. When selecting from locally relevant topics, an agent considers a number of recent posts by all agents they follow, and selects one at random. To determine the global importance and represent the range of event classes presented by Lehmann et al. we consider a family of functions that are defined by a linear increase and decrease around a peak of attention, with a minimum baseline outside of this period. By varying the beginning, peak and end of this shift in importance we are able to control the duration of each phase of the event and the rate of change observed therein. The behaviour of a authentic agent at time step $t$ is detailed in Algorithm~\ref{alg:legitimate_pseudocode}. An outline of the model is shown in Fig.~\ref{fig:model_schematic} and illustrates the selection of globally and locally important topics by an authentic agent.

In order to allow our model simulations to best reflect the real-world behaviour over social networks we use the Twitter egonet of followers available through the Stanford SNAP dataset~\cite{McAuley12}\footnote{https://snap.stanford.edu/data/ego-Twitter.html}. This directed network captures the small-world and scale-free properties frequently observed in online social systems, and includes 81,306 nodes and 1,768,149 edges. Our network representation considers directed edges $a \rightarrow b$ as indicating that $b$ follows $a$, i.e. posts are propagated in the direction of the edges. To leverage this directed nature, the local attention component of our model only considers those edges that point inwards.

For the following experiments we define a set of standard parameters outlined in Table~\ref{tab:def_params}. In order to produce a sensitivity analysis of these parameters, we vary them in turn. We define a standard functional form for the global importance of the simulated topic in Equation~\ref{eq:global_attention}. When varied this adjusts the start of the increase in attention (default step 125), the peak of attention (default step 250), the end of attention (default step 375), and the minimum attention (default 0.2).

\begin{table}
    \centering
    \begin{tabular}{|c|c|c|p{3cm}|c|c|c|}
        \hline Global awareness & Activity & Time steps & {\centering Prop. of spamming agents} & Memory & Spamming starts & Spamming ends \\ \hline
        0.4 &  0.3 & 500 & \centering{0} & 5 & 375 & 475 \\ \hline
    \end{tabular}
    \caption{The default parameters used in the agent-based model simulations. Global awareness determines the likelihood of selecting topic based on global importance, activity determines the likelihood of posting in a given time step, time steps determines the duration of the experiment, proportion of spamming agents determines the probability of an agent being defined to follow the spamming behaviour throughout the simulation, memory determines the number of past posts considered when selecting from topics used by neighbours and the spamming start and end values indicate the time step at which the spamming agents start and stop their activity.}
    \label{tab:def_params}
\end{table}

\begin{equation}
    f(t) = \left\{ \begin{array}{cc} 0.2 & \textrm{if } t \leq 125 \textrm{ or } t \geq 375 \\
    max(1-(250-t)/(250-125),0.2) & \textrm{if } 125<t<250 \\
    max(1-(t-250)/(375-250),0.2) & \textrm{otherwise.} \end{array}\right. \label{eq:global_attention}
\end{equation}

We ran a series of simulations to assess the impact that various model parameters have on the behaviour we observed, and detail them in Section~\ref{sec:varying model parameters}. The network size and duration of the simulation have little impact on the behaviour observed. Agent parameters similarly have limited impact once a sufficient level of engagement with the system is reached. The exception to this is the amount of spamming agents in the simulation, who can quickly dominate the properties seen in the scale-independent representation. Therefore, we believe that this model is representative of a range of social systems and we instead focus on how the scale-independent representation interacts with the properties of global attention.

\begin{figure}
    \centering
    \begin{subfigure}{0.45\textwidth}
    \includegraphics[width=\linewidth]{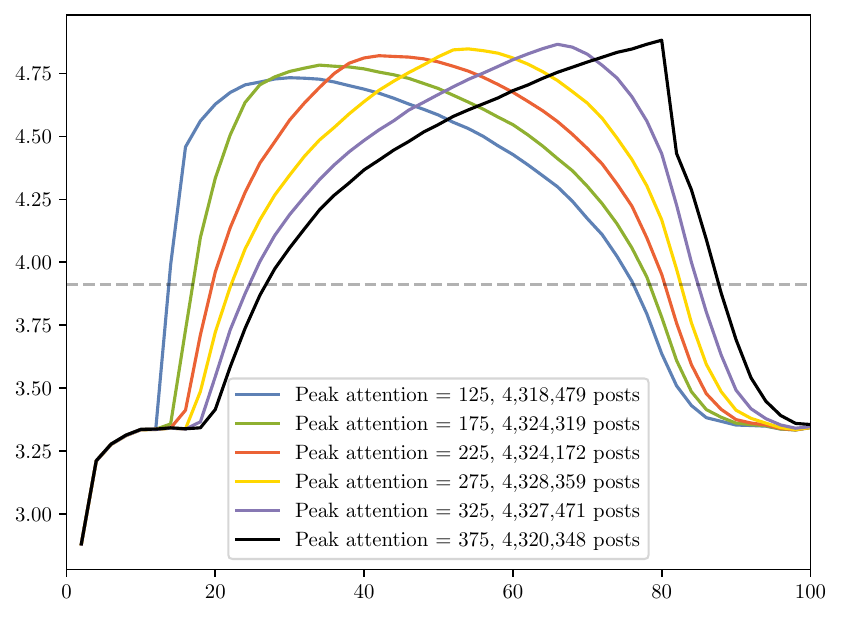}
    \caption{Varying the peak value, that is the time step of maximum global attention.}
    \label{fig:vary_peak}
    \end{subfigure}
    \hfill
    \begin{subfigure}{0.45\textwidth}
    \includegraphics[width=\linewidth]{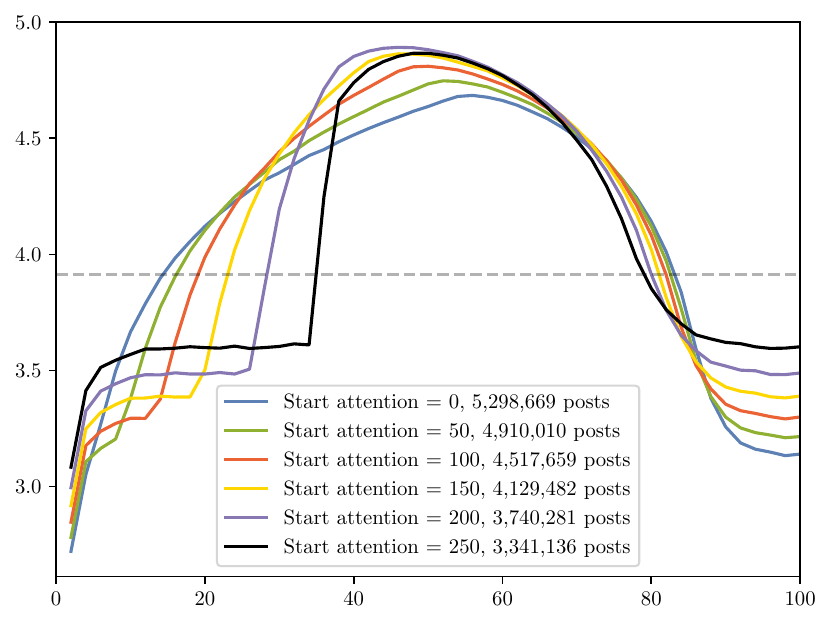}
    \caption{Varying the start value, that is the time step at which global attention begins to increase.}
    \label{fig:vary_start}
    \end{subfigure}
    \\
    \begin{subfigure}{0.45\textwidth}
    \includegraphics[width=\linewidth]{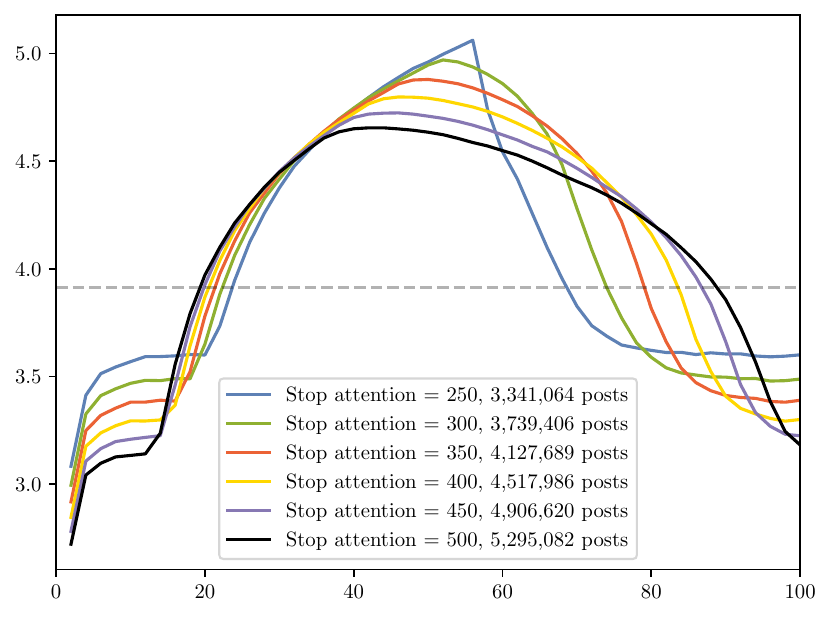}
    \caption{Varying the stop value, that is the time step at which global attention returns to baseline rate.}
    \label{fig:vary_stop}
    \end{subfigure}
    \hfill
    \begin{subfigure}{0.45\textwidth}
        \includegraphics[width=\linewidth]{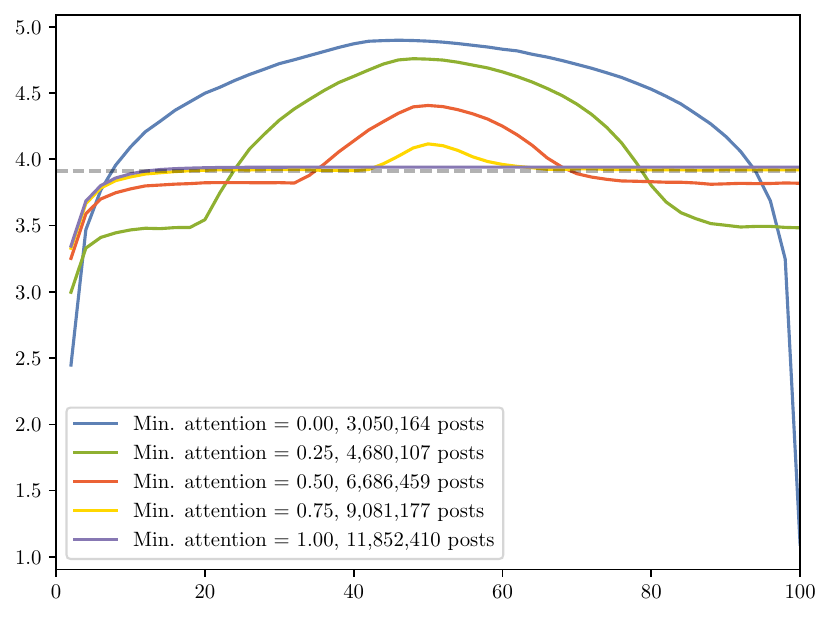}
        \caption{Varying the minimum value, that is the baseline level for global attention.}
        \label{fig:vary_min}
    \end{subfigure}
    \caption{The impact of running the model with different global attention functions, determining the importance of the simulated topic at the global level. We vary a) the peak time step, b) the start time step, c) the stop time step and d) the minimum level of global attention. The dashed line indicates the value of uniform, constant behaviour across the simulation.}
    \label{fig:vary_glob_att}
\end{figure}

We now seek to demonstrate the effects of varying the function describing the trends of global attention over time as given in Equation~\ref{eq:global_attention}. In Fig.~\ref{fig:vary_glob_att} we vary the start time, peak time, stop time and minimum value for this trend respectively. This figure further reinforces the observed trend in Fig.~\ref{sh:TSNE} that the patterns of collective attention form a continuum rather than a series of discrete categories. Fig.~\ref{fig:vary_peak} illustrates this best since varying only the time step of peak global attention is sufficient to transition from behaviour equivalent to the right-tailed (peak=125) through arch shaped (the default, peak=250) and into the left-tailed (peak=375). Through these changes we are able to recognise one possible type of behaviour that produces the sudden changes in topic engagement seen in the left- and right-tailed shapes of activity. As the peak time step in the global attention function approaches either the start or stop point of the activity interval, the sudden shift in the rate of topical posts reflects the sudden shift in global attention to the topic.

Considering now the changes in the start and stop times for the global attention in Figs.~\ref{fig:vary_start} and~\ref{fig:vary_stop} we see additional trends. Firstly, the changes in the scale-independent representation are not as sharp as in Fig.~\ref{fig:vary_peak}, except where the start and peak times or peak and stop times converge. This effect is likely due to the subsequent propagation of the topic of interest through social connections from early adopters. Comparing the total number of topical posts across the simulations shows an element of symmetry in the scale of attention received. Model runs with the same duration of increased global attention see a similar amount of posts regardless of the relative positions of the peak, whereas complementary adjustments to the start or stop times alone produce similar variation in the total number of posts.

The final aspect of the global attention function we seek to vary is the minimum value (set at the default of 0.2 in Equation~\ref{eq:global_attention}). In Fig.~\ref{fig:vary_min} we visualise the behaviour as this minimum changes. We can see that as the minimum value for global attention to the topic increases, the distinction between the event period and normal behaviour decreases. This is not surprising since increasing the minimum attention reduces the period within which the global attention function exceeds the (increased) minimum value. This case highlights another important aspect of the scale-independent representation, that is its ability to highlight the maximum change in activity, rather than the maximum volume of activity. The two extreme cases in Fig.~\ref{fig:vary_min} illustrate this perfectly along with their total post counts. When the minimum value of global attention is set to 0, we see fewer posts overall but a larger shift in attention patterns describing the arch-shaped behaviour. When the minimum value is set to 1, we instead see a much larger number of posts and consistent behaviour across much of the simulation, excluding the early period where there are fewer posts in the system to stimulate uptake of the topic through local factors.

The previous results have shown that our agent-based model is able to produce similar behaviour to that seen in Twitter, and reinforces the argument that instead of discrete classes we see a continuum of behaviours depending on the combination of factors influencing collective attention. One factor of the behaviour observed in Twitter that has not been recovered through the agent-based model is the size of the maximum values in the scale-independent representation. We believe this is a limitation of the discretisation of time in our model. Our model gives each user a single chance to post at a given time step and activates them in a random order, whereas in social systems users can interact with more variable frequency and post in rapid succession around topics of particular interest. This behaviour is likely necessary to produce the concentrated activity required for the scale-independent representation to return particularly large values. This is because the representation captures periods wherein the usage of a topic of interest is high relative to the usage of all topics. Despite this minor shortcoming in our modelling framework we have presented strong evidence that a few mechanistic factors can reproduce the qualitative behaviours of collective attention events.

\section{Discussion}\label{sec:discussion}

In this paper we presented a new method for studying the temporal trends of collective attention, with particular reference to how hashtag usage patterns change over periods of high interest. Our consideration of relative usage rates through the scale-independent representation overcomes two key limitations in large scale analysis of social media datasets by accounting for variation in both the total numbers of uses and the timeframe of interest, along with addressing the influence of exogenous drivers of activity such as time of day or longer-timescale trends. As a result, our method allows for a direct comparison of activity profiles across scales that were previously confounding. We supported this innovation through the use of an agent-based model and demonstrated that simple agent properties are sufficient to reproduce the activity profiles observed in social media data.

In Figs.~\ref{sh:fig:cpc18} and~\ref{sh:fig:esthermcvey} we showed that under previous timeseries techniques the choice of timescales strongly influences our perception of different events and can obscure processes that occur over different lengths of time. Through these examples we showed that the scale-independent representations reflect the underlying dynamics of these hashtags and present a non-parametric method which does not require any \emph{a priori} decision of which timescale to consider.  

Our comparison with the four classes found by Lehmann et al.~\cite{Lehmann12} showed that our method recovers the expected behaviour among common activity types. The ``before and during the peak'', ``during and after the peak'' and ``symmetrically around the peak'' classes from their work correspond to our left-tail, right-tail and arch-shaped profiles respectively, and demonstrate that they reflect an underlying process and are not an artefact of the choice of bin size in the timeseries. We did not find an analogue for the ``peak only'' class found by Lehmann et al. Intuitively, this is to be expected; under Lehmann et al.'s approach certain short-lived events fall entirely within a single day and as such are reduced to a timeseries with a single non-zero data point. In this sense, the ``peak only'' category is operating as a filter for event timescale. This is not an issue for our scale-independent methodology which can be applied to events occurring over a single day, or many days, without any changes to the process. 

A special note should be made about the abrupt shift shape characterised by hashtags such as \emph{amazon}, \emph{italexit}, \emph{putin} and \emph{trump}. This shape may look similar to Lehmann et al.'s ``peak only'' class but suggests more about the underlying system dynamics than the daily counts. These events demonstrate none of the gradual build up or decline in attention levels found under preferential attachment~\cite{Barabasi05} that should be considered characteristic of true collective attention. They instead represent a change of mode between a baseline activity state and a more rapid event state. Such bimodal behaviour is unlikely to occur naturally in a self-organising system and suggests that certain exogenous perturbations are applied. After examining specimen tweets from these hashtags there is reason to believe that this shape is characteristic of astroturfing and other artificial attempts to boost content. In the case of hashtags like \emph{amazon} and \emph{italexit}, tweets during these periods of activity frequently include variations on a comment that is trying to be publicised alongside a number of hashtags to promote visibility. This is a common tactic with spamming behaviour, and will usually include several of the currently trending hashtags to promote visibility in as many searches as possible~\cite{Lee12}. The activation of automated accounts would also help to explain the suddenness of the transition between modes in the hashtag profile and the stark contrast from the more typical behaviour of more gradual transitions on one or both sides of the peak.

Through our agent-based model simulations we defined mechanisms which produced the same activity patterns as described in Section~\ref{sec:shapes} and are complementary to the original classes first described by Lehmann et al. We found further evidence to suggest that these patterns do not constitute distinct classes but instead form a continuum of behaviours once the requirement to choose a timescale is resolved. We also saw that, as anticipated from the Twitter examples, the abrupt shift activity profile is reliably reproduced when even a small proportion of spamming agents are included in the model. Moreover, the range of global importance functions considered in Fig.~\ref{fig:vary_glob_att} demonstrated that the key characteristic of the abrupt shift class (that is the sudden shift to and from the event mode) is likely only produced by sudden shifts in the relative popularity of a given topic, as opposed to the social growth of collective attention.

Following these observations, we should also highlight how the shapes observed in our scale-independent representation can be used to infer generating mechanisms more widely. Considering parts of the shape based on how their representation changes, we recall that large changes are necessarily linked to periods where the activity rate of a given hashtag changes rapidly compared to the background rate. During the emergence of an event, such a sudden increase suggests that some external stimulus or rapid viral growth is driving new attention, whereas a sudden decline shows that the hashtag quickly loses relevance after its peak. Smaller changes, however, are indicative of the preferential attachment-like behaviour required for our preferred definition of collective attention. Interpretation of the dynamics governing the following decline in attention is more challenging, but a gradual decline could arise from events which remain relevant beyond the initial peak through repeated updates or emergent conversations. These links with the underlying dynamics, and preferential attachment in particular, are an important advantage of our new method over existing timeseries techniques.

When taking the agent-based model simulations into consideration we can make additional inferences between topic circumstances and the type of behaviour we observe. Our sensitivity analysis of the scale-independent representation of activity under different parameter settings showed that much of the behaviour we saw in the Twitter dataset is likely to be driven by variations in the global attention to a given topic. While simulations did show that the social diffusion of topics can affect the usage patterns, these features were relatively minor when compared to those seen through variation of the global attention function.

Despite our characterisation of four notable profiles, Figs.~\ref{sh:TSNE} and~\ref{fig:vary_glob_att} show that these shapes do not fall within distinct categories but instead form points on a continuum. In light of this we argue that the characteristic shapes we highlight above should be used to help infer the underlying behaviour of collective attention, rather than labelling a particular activity period. Tails in the hashtag profile suggest gradual changes in activity, whereas sudden increases signify rapid changes. Understanding these features of the hashtag profiles may allow scholars to infer the mechanistic behaviour of the hashtag without reliance on categories which would be poorly fit to most of our dataset.


In light of the results in the preceding sections, we recommend that future analysis of behavioural trends in online social media should avoid methods which bin to coarse resolution, since social media interactions can be too intense and fleeting to be properly captured with this approach. Our scale-independent representation permits comparisons that are not possible under conventional approaches such as natural disasters in significantly different contexts. Given a period of interest, applying this methodology will provide indications of the behaviour within, and suggest whether multiple distinct events are captured. Subsequent subdivision of the period can be used in such cases to concentrate on specific intervals, and we demonstrate an example of this using \#cpc18 in Supplementary Information Section~\ref{sec:splitting-cpc18}. The agent-based model we describe here can then be used to support inferences about how the collective attention patterns emerge through approximation of the scale-independent representation.

The existence of a continuum of collective attention profiles rather than distinct classes suggests that future attempts to develop automated classification tools for behavioural types will be difficult and potentially misleading in terms of the relationships between these types. Efforts treating the scale-independent representation as a vector will likely encounter the same challenges seen in Fig.~\ref{sh:TSNE} and may succeed in separating left- and right-tailed shapes, whereas the other prototypical behaviours will prove more difficult. Automated fitting of the agent-based model to estimate parameters may be more tractable but will encounter additional problems in reproducing the maximum amplitude and efficiently exploring the expansive space of possible global attention functions.



There is additional scope to extend the work presented here into other application domains. While we designed the scale-independent representation and agent-based model to be adaptable to a wide range of social systems, our tests here only assess the case of online collective attention. We expect that the four prototypical shapes of collective attention cover all types of human behaviour, but it remains to be seen how common each of these behaviours are in social systems away from social media. Future work should investigate these applications.

Our analyses here demonstrate that while collective attention events may occur over systems of different scales, in many cases they are driven by the similar patterns in human behaviour. The new tools we introduce here should allow scholars to gain additional perspectives across a wide range of domains and moreover highlight domain specific effects on collective attention. We look forward to further work in this area and deeper knowledge of the universality of collective attention processes.

\subsection*{Acknowledgements}
The authors thank Veronica White for design expertise for Fig.~\ref{fig:model_schematic}. \\

\noindent For the purpose of open access, the author has applied a Creative Commons Attribution (CC BY) licence to any Author Accepted Manuscript version arising from this submission.

\subsection*{Author contributions}
TJBC and ISW developed the scale-independent representation. ISW collected the Twitter data. TJBC developed the agent-based model, carried out all data analysis and wrote the first draft of the manuscript. All authors reviewed and approved the final version of the manuscript.

\subsection*{Funding}
TJBC is based at the Centre for Climate Communication and Data Science which is part funded by the University of Exeter and Children’s Investment Fund Foundation (CIFF) (grant number 2210-08101). Parts of this work were completed during an EPSRC PhD studentship (grant number
EP/M506527/1). The funders had no role in the conceptualization, design, data collection, analysis, decision to publish, or preparation of the manuscript and therefore the findings and conclusions are those of the authors and do not necessarily reflect the positions or policies of the funders.

\bibliographystyle{abbrv}
\bibliography{refs}

\begin{thebibliography}{10}

\bibitem{Akbarpour18}
M.~Akbarpour and M.~O. Jackson.
\newblock Diffusion in networks and the virtue of burstiness.
\newblock {\em Proceedings of the National Academy of Sciences},
  115(30):E6996--E7004, 2018.

\bibitem{Barabasi05}
A.-L. Barab{\'a}si.
\newblock The origin of bursts and heavy tails in human dynamics.
\newblock {\em Nature}, 435(7039):207--211, 2005.

\bibitem{Barabasi99}
A.-L. Barab{\'a}si and R.~Albert.
\newblock Emergence of scaling in random networks.
\newblock {\em Science}, 286(5439):509--512, 1999.

\bibitem{Bathina17}
K.~C. Bathina, A.~Jammalamadaka, J.~Xu, and T.-C. Lu.
\newblock An agent-based model of posting behavior during times of societal
  unrest.
\newblock In {\em Social, Cultural, and Behavioral Modeling}, pages 53--59,
  2017.

\bibitem{Boulton19}
C.~Boulton and T.~Lenton.
\newblock A new method for detecting abrupt shifts in time series.
\newblock {\em F1000Research}, 8(746), 2019.

\bibitem{Burnap14}
P.~Burnap, M.~L. Williams, L.~Sloan, O.~Rana, W.~Housley, A.~Edwards,
  V.~Knight, R.~Procter, and A.~Voss.
\newblock Tweeting the terror: modelling the social media reaction to the
  {W}oolwich terrorist attack.
\newblock {\em Social Network Analysis and Mining}, 4(1):206, 2014.

\bibitem{Cann21}
T.~J.~B. Cann, I.~S. Weaver, and H.~T.~P. Williams.
\newblock Ideological biases in social sharing of online information about
  climate change.
\newblock {\em PLOS ONE}, 16(4):1--25, 2021.

\bibitem{Castillo14}
C.~Castillo, M.~El-Haddad, J.~Pfeffer, and M.~Stempeck.
\newblock Characterizing the life cycle of online news stories using social
  media reactions.
\newblock In {\em Proceedings of the 17th ACM Conference on Computer Supported
  Cooperative Work \& Social Computing}, CSCW ’14, page 211–223, 2014.

\bibitem{DeDomenico20}
M.~De~Domenico and E.~G. Altmann.
\newblock Unraveling the origin of social bursts in collective attention.
\newblock {\em Scientific Reports}, 10(1):4629, 2020.

\bibitem{Ertugrul19}
A.~M. Ertugrul, Y.-R. Lin, W.-T. Chung, M.~Yan, and A.~Li.
\newblock Activism via attention: interpretable spatiotemporal learning to
  forecast protest activities.
\newblock {\em EPJ Data Science}, 8(1):5, 2019.

\bibitem{deArruda17}
G.~Ferraz~de Arruda, F.~Aparecido~Rodrigues, P.~Martín~Rodríguez, E.~Cozzo,
  and Y.~Moreno.
\newblock {A general Markov chain approach for disease and rumour spreading in
  complex networks}.
\newblock {\em Journal of Complex Networks}, 6(2):215--242, 2017.

\bibitem{Garimella17}
K.~Garimella, G.~De~Francisci~Morales, A.~Gionis, and M.~Mathioudakis.
\newblock The effect of collective attention on controversial debates on social
  media.
\newblock In {\em Proceedings of the 2017 ACM on Web Science Conference},
  WebSci '17, page 43–52, 2017.

\bibitem{Gleeson14}
J.~P. Gleeson, D.~Cellai, J.-P. Onnela, M.~A. Porter, and F.~Reed-Tsochas.
\newblock A simple generative model of collective online behavior.
\newblock {\em Proceedings of the National Academy of Sciences},
  111(29):10411--10415, 2014.

\bibitem{Harush17}
U.~Harush and B.~Barzel.
\newblock Dynamic patterns of information flow in complex networks.
\newblock {\em Nature Communications}, 8(1):2181, 2017.

\bibitem{He17}
X.~He and Y.-R. Lin.
\newblock Measuring and monitoring collective attention during shocking events.
\newblock {\em EPJ Data Science}, 6(1):30, 2017.

\bibitem{Healy15}
P.~{Healy}, G.~{Hunt}, S.~{Kilroy}, T.~{Lynn}, J.~P. {Morrison}, and
  S.~{Venkatagiri}.
\newblock Evaluation of peak detection algorithms for social media event
  detection.
\newblock In {\em 2015 10th International Workshop on Semantic and Social Media
  Adaptation and Personalization (SMAP)}, pages 1--9, 2015.

\bibitem{Huynh15}
H.~N. Huynh, E.~F. Legara, and C.~Monterola.
\newblock A dynamical model of {T}witter activity profiles.
\newblock In {\em Proceedings of the 26th ACM Conference on Hypertext \& Social
  Media}, HT ’15, page 49–57, 2015.

\bibitem{Kitsak10}
M.~Kitsak, L.~K. Gallos, S.~Havlin, F.~Liljeros, L.~Muchnik, H.~E. Stanley, and
  H.~A. Makse.
\newblock Identification of influential spreaders in complex networks.
\newblock {\em Nature Physics}, 6(11):888--893, 2010.

\bibitem{Lee12}
K.~Lee, J.~Caverlee, K.~Y. Kamath, and Z.~Cheng.
\newblock Detecting collective attention spam.
\newblock In {\em Proceedings of the 2nd Joint WICOW/AIRWeb Workshop on Web
  Quality}, WebQuality ’12, page 48–55, 2012.

\bibitem{Lehmann12}
J.~Lehmann, B.~Gon\c{c}alves, J.~J. Ramasco, and C.~Cattuto.
\newblock Dynamical classes of collective attention in {T}witter.
\newblock In {\em Proceedings of the 21st International Conference on World
  Wide Web}, WWW '12, page 251–260, 2012.

\bibitem{Lin14}
Y.-R. Lin, B.~Keegan, D.~Margolin, and D.~Lazer.
\newblock Rising tides or rising stars?: Dynamics of shared attention on
  {T}witter during media events.
\newblock {\em PLOS ONE}, 9(5):1--12, 2014.

\bibitem{Lin13}
Y.-R. Lin, D.~Margolin, B.~Keegan, A.~Baronchelli, and D.~Lazer.
\newblock \#{B}igbirds never die: Understanding social dynamics of emergent
  hashtags.
\newblock {\em Seventh International AAAI Conference on Weblogs and Social
  Media}, 2013.

\bibitem{LorenzSpreen18}
P.~Lorenz-Spreen, F.~Wolf, J.~Braun, G.~Ghoshal, N.~Djurdjevac~Conrad, and
  P.~H{\"o}vel.
\newblock Tracking online topics over time: understanding dynamic hashtag
  communities.
\newblock {\em Computational Social Networks}, 5(1):9, 2018.

\bibitem{McAuley12}
J.~McAuley and J.~Leskovec.
\newblock Learning to discover social circles in ego networks.
\newblock In {\em Proceedings of the 25th International Conference on Neural
  Information Processing Systems - Volume 1}, NIPS'12, page 539–547, Red
  Hook, NY, USA, 2012. Curran Associates Inc.

\bibitem{Medvedev19}
A.~N. Medvedev, J.-C. Delvenne, and R.~Lambiotte.
\newblock {Modelling structure and predicting dynamics of discussion threads in
  online boards}.
\newblock {\em Journal of Complex Networks}, 7(1):67--82, 2018.

\bibitem{Mitra17}
T.~Mitra, G.~Wright, and E.~Gilbert.
\newblock Credibility and the dynamics of collective attention.
\newblock {\em Proc. ACM Hum.-Comput. Interact.}, 1(CSCW), 2017.

\bibitem{Moutidis19}
I.~Moutidis and H.~T.~P. Williams.
\newblock Utilizing complex networks for event detection in heterogeneous
  high-volume news streams.
\newblock In {\em Complex Networks and Their Applications VIII}, pages
  659--672, 2020.

\bibitem{Olteanu15}
A.~Olteanu, C.~Castillo, N.~Diakopoulos, and K.~Aberer.
\newblock Comparing events coverage in online news and social media: The case
  of climate change.
\newblock In {\em International AAAI Conference on Web and Social Media}, 2015.

\bibitem{Saito15}
S.~Saito, Y.~Hirata, K.~Sasahara, and H.~Suzuki.
\newblock Tracking time evolution of collective attention clusters in
  {T}witter: Time evolving nonnegative matrix factorisation.
\newblock {\em PLOS ONE}, 10(9):1--17, 2015.

\bibitem{Sasahara13}
K.~Sasahara, Y.~Hirata, M.~Toyoda, M.~Kitsuregawa, and K.~Aihara.
\newblock Quantifying collective attention from tweet stream.
\newblock {\em PLOS ONE}, 8(4):1--10, 2013.

\bibitem{Segerberg11}
A.~Segerberg and W.~L. Bennett.
\newblock Social media and the organization of collective action: Using
  {T}witter to explore the ecologies of two climate change protests.
\newblock {\em The Communication Review}, 14(3):197--215, 2011.

\bibitem{Starbird12}
K.~Starbird and L.~Palen.
\newblock ({H}ow) will the revolution be retweeted? {I}nformation diffusion and
  the 2011 {E}gyptian uprising.
\newblock In {\em Proceedings of the ACM 2012 Conference on Computer Supported
  Cooperative Work}, CSCW ’12, page 7–16, 2012.

\bibitem{tenThij19}
M.~{ten Thij}, A.~Kaltenbrunner, D.~Laniado, and Y.~Volkovich.
\newblock Collective attention patterns under controlled conditions.
\newblock {\em Online Social Networks and Media}, 13:100047, 2019.

\bibitem{vanderMaaten08}
L.~van~der Maaten and G.~Hinton.
\newblock Visualizing high-dimensional data using {t-SNE}.
\newblock {\em Journal of Machine Learning Research}, 9:2579--2605, 2008.

\bibitem{Vosoughi18}
S.~Vosoughi, D.~Roy, and S.~Aral.
\newblock The spread of true and false news online.
\newblock {\em Science}, 359(6380):1146--1151, 2018.

\bibitem{Wang19}
L.-Z. Wang, Z.-D. Zhao, J.~Jiang, B.-H. Guo, X.~Wang, Z.-G. Huang, and Y.-C.
  Lai.
\newblock A model for meme popularity growth in social networking systems based
  on biological principle and human interest dynamics.
\newblock {\em Chaos: An Interdisciplinary Journal of Nonlinear Science},
  29(2):023136, 2019.

\bibitem{Weng12}
L.~Weng, A.~Flammini, A.~Vespignani, and F.~Menczer.
\newblock Competition among memes in a world with limited attention.
\newblock {\em Scientific Reports}, 2(1):335, 2012.

\bibitem{Weng14}
L.~Weng, F.~Menczer, and Y.-Y. Ahn.
\newblock Predicting successful memes using network and community structure.
\newblock In {\em International AAAI Conference on Web and Social Media}, 2014.

\bibitem{Wu07}
F.~Wu and B.~A. Huberman.
\newblock Novelty and collective attention.
\newblock {\em Proceedings of the National Academy of Sciences},
  104(45):17599--17601, 2007.

\bibitem{Yang12}
L.~Yang, T.~Sun, M.~Zhang, and Q.~Mei.
\newblock We know what \@you \#tag: Does the dual role affect hashtag adoption?
\newblock In {\em Proceedings of the 21st International Conference on World
  Wide Web}, WWW ’12, page 261–270, 2012.

\bibitem{Zhao20}
Z.~Zhao, J.~Zhao, Y.~Sano, O.~Levy, H.~Takayasu, M.~Takayasu, D.~Li, J.~Wu, and
  S.~Havlin.
\newblock Fake news propagates differently from real news even at early stages
  of spreading.
\newblock {\em EPJ Data Science}, 9(1):7, 2020.

\end{thebibliography}

\appendix

\pagebreak

\section*{Supplementary information}

\renewcommand{\thefigure}{S\arabic{figure}}
\renewcommand{\thetable}{S\arabic{table}}
\renewcommand{\thealgorithm}{S\arabic{algorithm}}
\renewcommand{\thesection}{S\arabic{section}}

\setcounter{figure}{0}
\setcounter{table}{0}

\section{Illustrations of previously identified classes of collective attention}

\begin{figure}[h]
    \centering
    \begin{subfigure}{0.35\textwidth}
    \includegraphics[width=\linewidth]{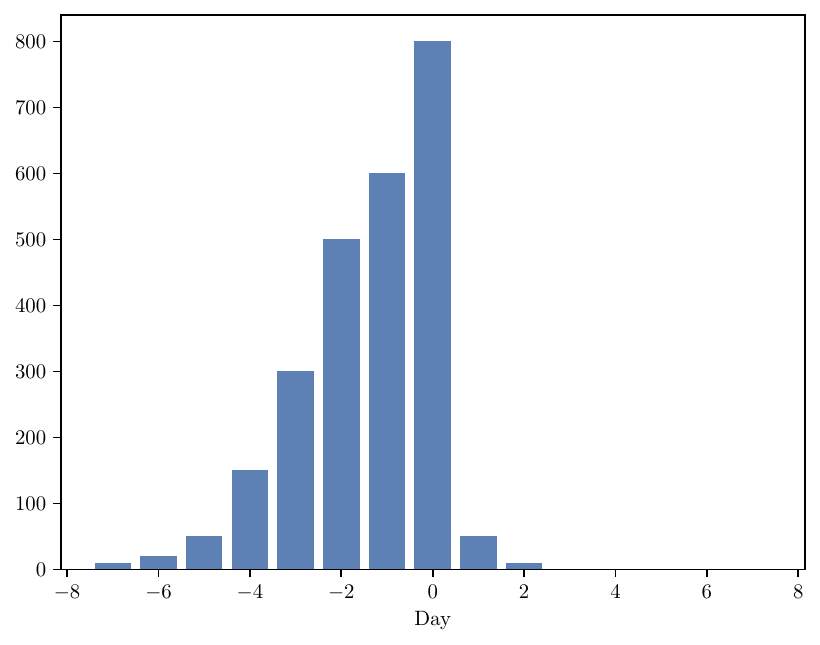}
    \caption{``Before the peak'' class}\label{sh:fig:eg_before}
    \end{subfigure} 
    \hfill
    \begin{subfigure}{0.35\textwidth}
    \includegraphics[width=\linewidth]{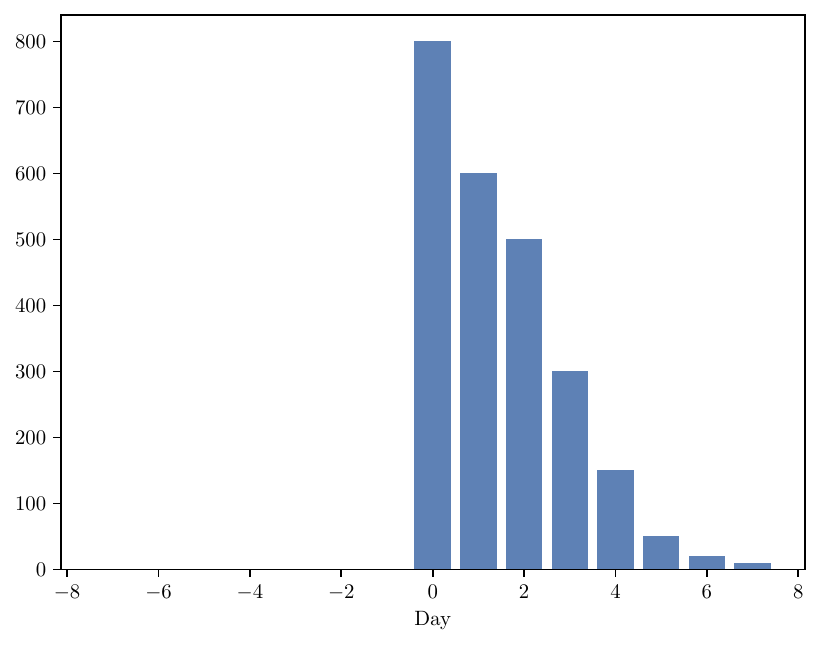}
    \caption{``After the peak'' class}\label{sh:fig:eg_after}
    \end{subfigure} \\
    \begin{subfigure}{0.35\textwidth}
    \includegraphics[width=\linewidth]{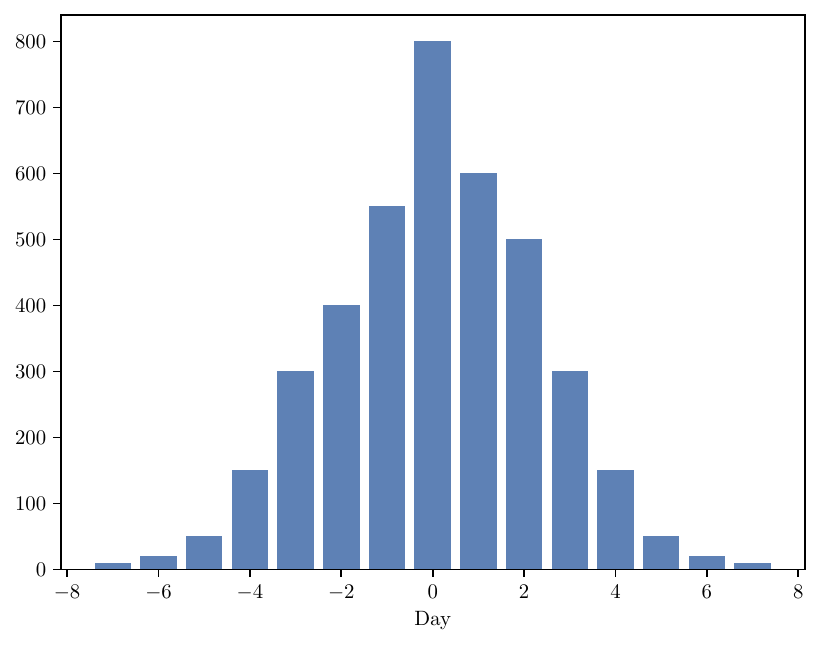}
    \caption{``Before and after the peak'' class}\label{sh:fig:eg_both}
    \end{subfigure} 
    \hfill
    \begin{subfigure}{0.35\textwidth}
    \includegraphics[width=\linewidth]{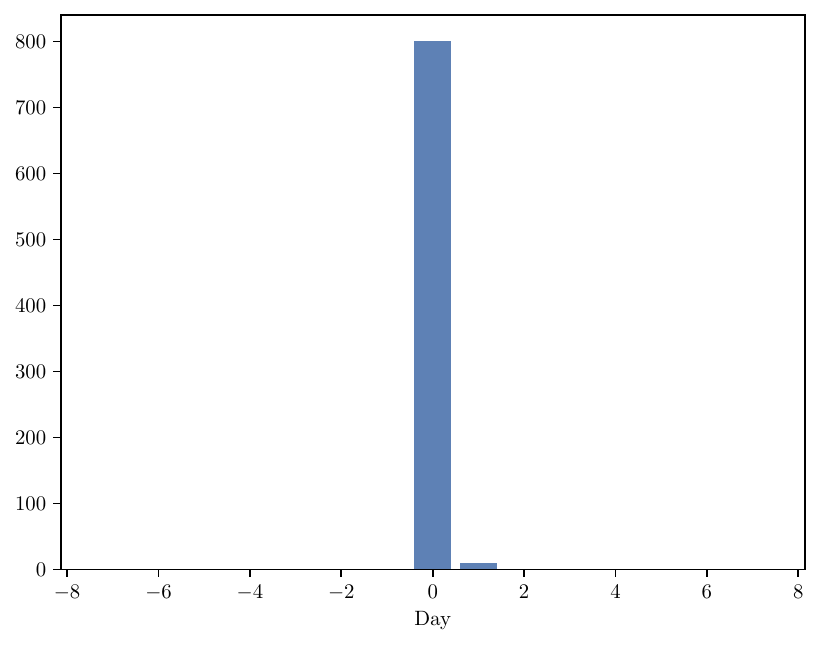}
    \caption{``Peak only'' class}\label{sh:fig:eg_peak}
    \end{subfigure}
    \caption{Illustrations of the activity profiles represented by each of the four classes found by Lehmann et al.~\cite{Lehmann12}. Each illustration is centered on day 0, the day of peak activity.}
    \label{sh:fig:eg_classes}
\end{figure}

\pagebreak
\section{Partitioning hashtag usage in the dataset}

\begin{figure}[h]
    \centering
    \includegraphics[width=\textwidth]{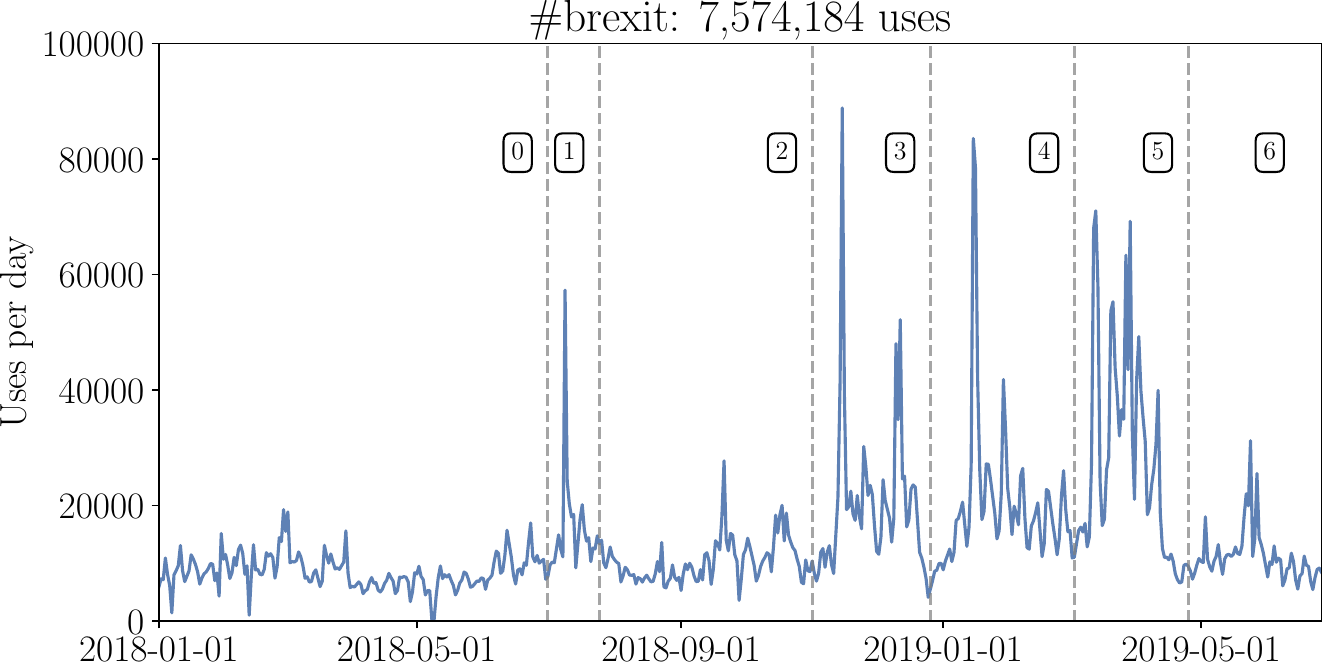}
    \caption{Each hashtag is divided into a series of intervals over the study period based on periods of increased activity. In this example, seven periods are defined. The numeric labels for each period are indicative of the interval numbers as referenced throughout this manuscript.}
    \label{sh:brexit splits}
\end{figure}

Fig.~\ref{sh:brexit splits} illustrates the partitioning process for activity over the whole dataset, isolating periods of increased activity (i.e. intervals 1,3,4,5) from periods of variability within a normal range (i.e. intervals 0,2,6). 

\pagebreak
\section{The impact of timescale on the ``peak only'' class}

\begin{figure}[h]
    \centering
    \begin{subfigure}{0.45\textwidth}
    \includegraphics[width=\linewidth]{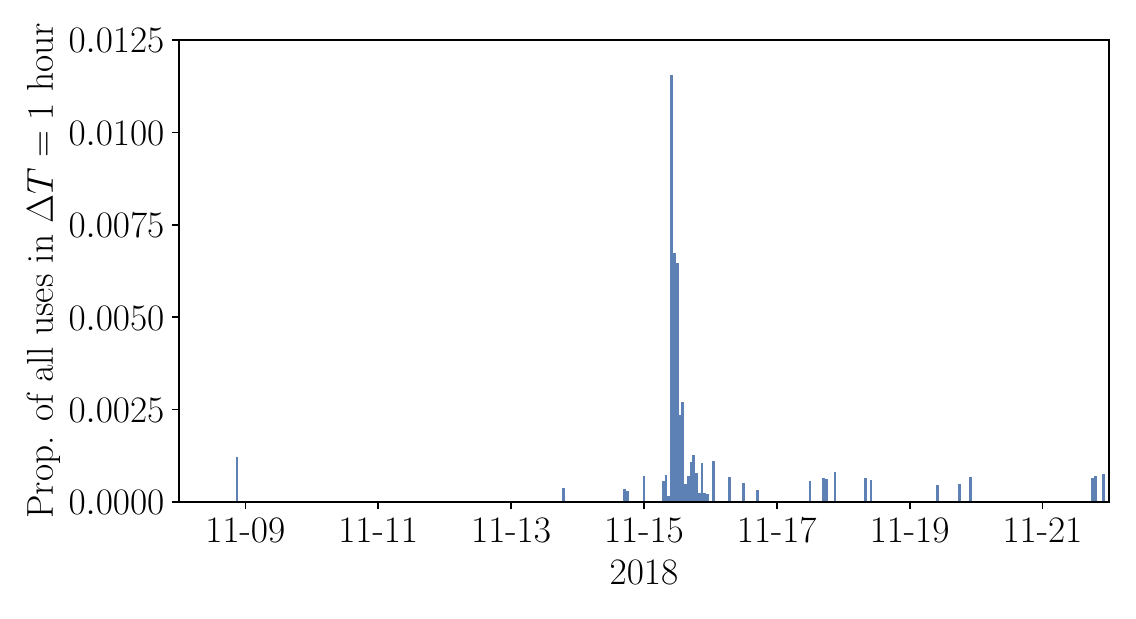}
    \caption{Timeseries with one-hour bin width.}\label{sh:fig:esthermcvey_a}
    \end{subfigure}
    \hfill
    \begin{subfigure}{0.45\textwidth}
    \includegraphics[width=\linewidth]{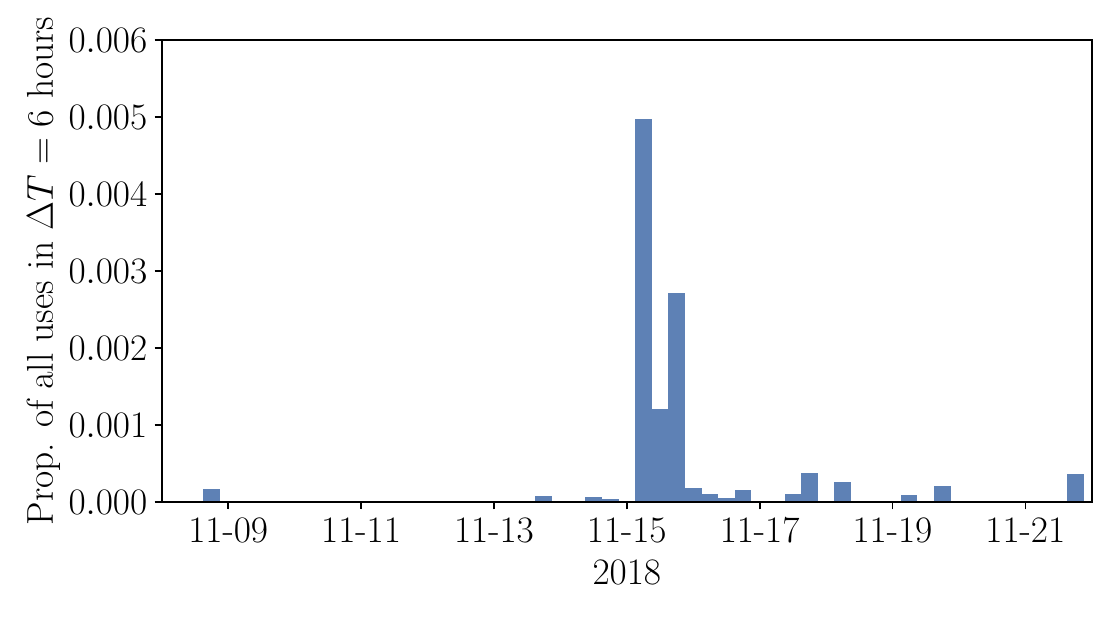}
    \caption{Timeseries with six-hour bin width.}\label{sh:fig:esthermcvey_b}
    \end{subfigure} \\
    \begin{subfigure}[t]{0.45\textwidth}
    \includegraphics[width=\linewidth]{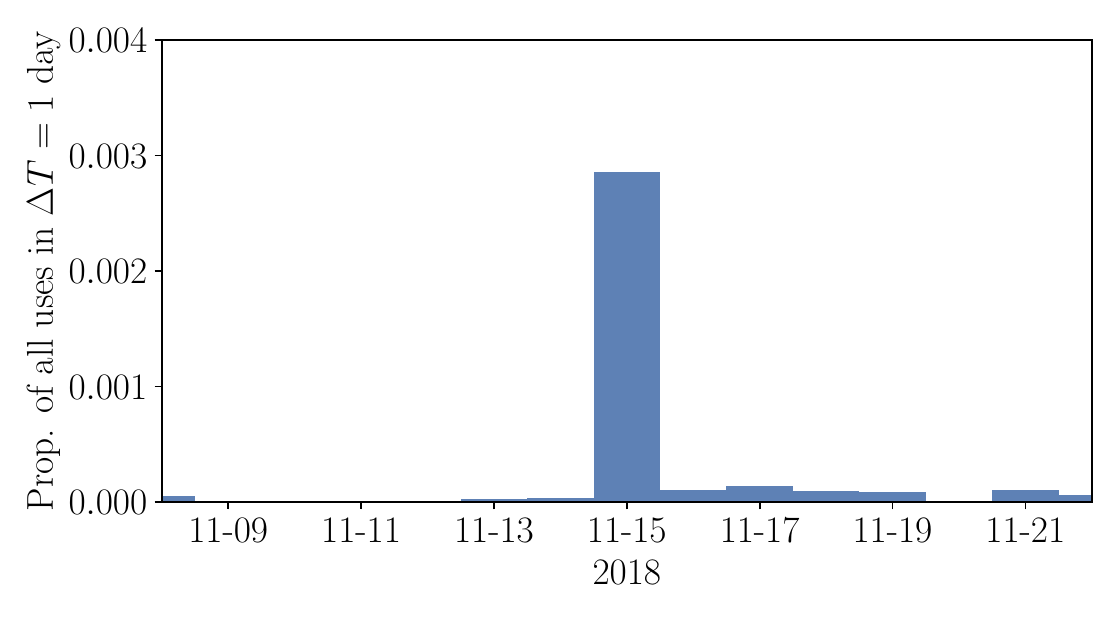}
    \caption{Timeseries with one-day bin width.}\label{sh:fig:esthermcvey_c}
    \end{subfigure}
    \hfill
    \begin{subfigure}[t]{0.45\textwidth}
    \includegraphics[width=\linewidth]{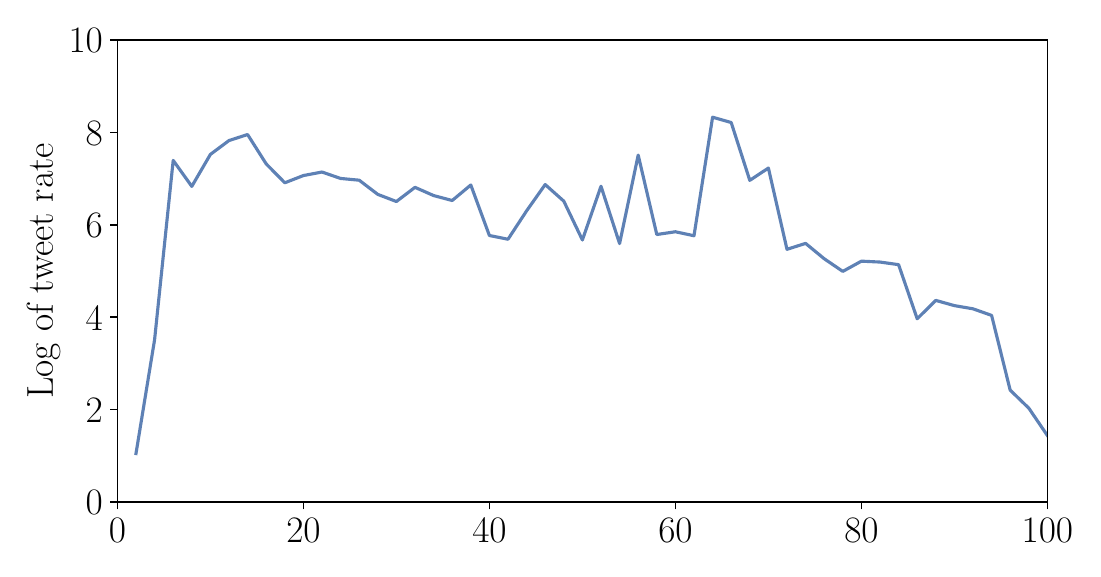}
    \caption{Scale-independent representation using the hashtag profile with $N=50$.}\label{sh:fig:esthermcvey_d}
    \end{subfigure}
    \caption{\emph{\#esthermcvey}, referring to British MP Esther McVey and her resignation as a cabinet minister. Despite most of the use of this hashtag occurring on the peak day, there is evidence of a gradual decline on shorter timescales.}
    \label{sh:fig:esthermcvey}
\end{figure}

Here we look further at the peak-only class proposed by Lehmann et al~\cite{Lehmann12}. Our categorisation of this class only considers events for which the peak day accounts for $90\%$ of usage activity for the hashtag. Fig.~\ref{sh:fig:esthermcvey} shows one such hashtag \emph{\#esthermcvey}, in reference to the British MP Esther McVey who resigned as a cabinet minister on $15$ November $2018$ over opposition to a proposed Brexit deal. Social media attention to this event is short-lived, with most related tweets falling within a single day. With a smaller timescale it becomes apparent that this event is not characterised by a spontaneous end in the same way as its spontaneous start. The hourly resolution shows that interest tails off more gradually over a number of hours. The scale-independent representation captures this behaviour, showing a decline in activity from the peak despite the majority of the tweets occurring over the course of a couple of hours.

\pagebreak
\section{Selected examples of tweets using hashtags identified with an abrupt shift profile}

\begin{table}[h]
    \centering
    \begin{tabular}{|p{0.95\textwidth}|}
        \hline
         [...] Blow the whistle [...] \#wikileaks \#eagles \#nfl  \#amazon \#netflix \#iTunes \#Brexit \#nyc  \#starbucks \#uber \#hiphop \#music [...] lost even with DNC VoterFraud DOJ FBI Fake News Media  Shadow Banning and Russia Dossier. Another Surprise waiting in 2018 ? [...]  \\ \hline
         \#STOPCLANDESTINI \#STOPINVASIONE \#StopIslam \#stopong \#tolleranzazero \#chiudiamoiporti \#portichiusi \#blocconavalesubito \#iostoconsalvini \#NessunoTocchiSalvini \#italexit \#frexit \#grexit \#nexit \#brexit \#stopEU \#leaveEU \#Stopsoros \#SALVININONMOLLARE \\ \hline
         The \#Leave Campaign, aided by \#Putin, used unfettered capitalism during the \#BREXIT/\#LEXIT Referendum @JohnMcDonnellMP [...]Why \#Labour is @JeremyCorbyn helping @Conservatives to make Rupert Murdoch's day? \#StopBrexit \#WATON \#ABTV \#FBPE [...] \\
         \hline
         \#Anonymous [...] \#DAX \#Money \#Investing \#Wealth \#HNW \#UHNW \#Verm[...]gensverwaltung \#roboadvisor \#Brexit \#NationalSiblingsDay  [...] \\
         \hline
    \end{tabular}
    \caption{Selected examples of repeated tweet text during periods with an abrupt shift hashtag profile. For readability and anonymity, strings of unicode characters, URLs and private user mentions have been replaced with ``[...]''.}
    \label{sh:tab:shift examples}
\end{table}

\pagebreak
\section{Additional details for the agent-based model}

Here we provide pseudocode for the behaviour of the agents included in our simulation model. Two types of agents exist: spamming agents and authentic agents.

Spamming agents are covered in Algorithm~\ref{alg:botnet_pseudocode} and are designed to simulate the behaviour of a group of coordinated users flooding the communications system with a large number of posts about the given topic. These users are in one of two states, determined by the time step. In their passive mode, they do not post at all. In their active range (occurring between two given time steps) they constantly post on their chosen topic with no deviation in frequency or topic.

\begin{algorithm}
\caption{Step behaviour for spamming agents.}
\label{alg:botnet_pseudocode}
\begin{algorithmic}
\State $start \gets a$, $end \gets b$
\For{time step $t$}
\If{$a<=t<=b$}
\State $post \gets 2$
\Else
\State $post \gets 0$
\EndIf
\State Record $post$
\EndFor
\end{algorithmic}
\end{algorithm}

Authentic agents are covered in Algorithm~\ref{alg:legitimate_pseudocode}. These agents are designed to reflect expected behaviour of individuals who converse on different topics depending on those that occur around them. At each time step, an agent decides whether to post at all based on their personal activity rate. Should they choose to post, the topic of the post is determined by their balance of preference between content that is globally interesting (imagine that which appears in the trending feed of a social media platform) or relevant to their friends. We model the importance of our topic of interest as a family of functions of time to reflect different ways in which interest evolves. When selecting from the topics of interest to their friends however, the agent reviews a given number of recent posts by each person they follow before deciding to post on- or off-topic based on the balance therein.

\begin{algorithm}
\caption{Step behaviour for authentic agents. Note that the selection based on neighbour activity is dependent on activation order of agents at a given time step, but repeated simulation under additional random seeds show that the system-level behaviour observed is consistent.}
\label{alg:legitimate_pseudocode}
\begin{algorithmic}
\State Activity $\gets a$, global awareness $\gets gl\_aw$, global attention $\gets g(t)$, memory $\gets m$
\For{time step $t$}
\If{$Rand<a$} \Comment{Determine whether to post}
    \If{$Rand<gl\_aw$} \Comment{Choose based on global attention}
        \If{$Rand<g(t)$}
        \State $post \gets 2$
        \Else
        \State $post \gets 1$
        \EndIf
    \Else \Comment{Choose based on neighbour activity}
    \State Collect the most recent $m$ posts from each leader
    \State Choose post $p$ to be on- or off-topic based on the activity of the leaders:
    \State $Pr(p=2)=|\{$Leader posts $p=2\}|/|\{$Leader posts $p=1$ or $p=2$\}$|$ 
    \State $post \gets p$
    \EndIf
\Else
\State $post \gets 0$
\EndIf
\State Record $post$
\EndFor
\end{algorithmic}
\end{algorithm}

\pagebreak
\section{Varying model parameters}\label{sec:varying model parameters}

\begin{figure}[h]
    \centering
    \begin{subfigure}{0.45 \textwidth}
        \includegraphics[width=\linewidth]{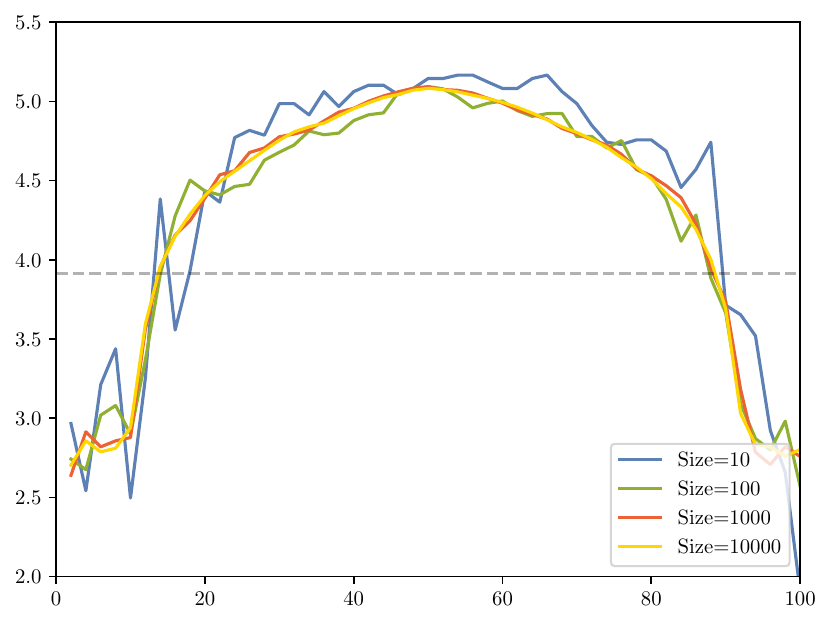}
        \caption{Tests varying the network size.}
        \label{supp_fig:testing_network_size}
    \end{subfigure} \hfill
    \begin{subfigure}{0.45 \textwidth}
        \includegraphics[width=\linewidth]{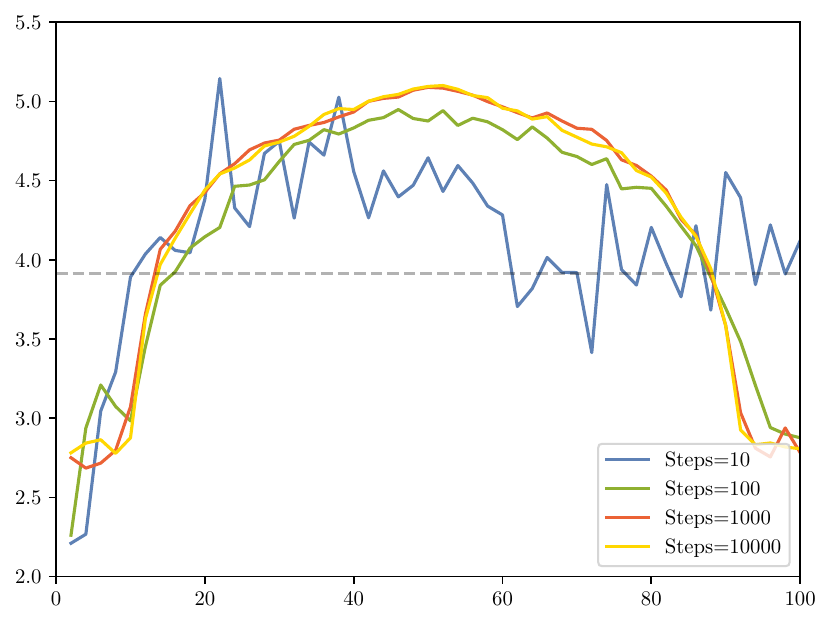}
        \caption{Tests varying the number of time steps.}
        \label{supp_fig:testing_time_steps}
    \end{subfigure}
    \caption{Varying scale in terms of the number of nodes and number of time steps does not affect the behaviour of the model. We run a series of simulations with Barabasi-Albert random networks with 500 nodes, each adding one edge; each agent has an activity rate 0.8, global awareness of 0.3 and no bots are included. We simulate the global preference towards the topic of interest with a function $f$ with minimum value 0.1, increasing linearly from 25\% of the steps to peak at 1 at 50\% of the steps before linearly decreasing to a minimum at 75\% of the steps.}
    \label{supp_fig:scale_variation}
\end{figure}

Here we report on the effect of varying the network or agent parameters in our model. Using simulation results from this agent-based model we are able to present additional justification that our proposed representation accurately reflects underlying behaviours irrespective of the number of users in the system and the duration of an event. We ran simulations on random networks generated from the Barabasi-Albert preferential attachment model~\cite{Barabasi99}, varying the size of the network and the number of time steps simulated while fixing attention parameters. Fig.~\ref{supp_fig:scale_variation} illustrates that once a sufficient scale has been reached to minimise the impact of local noise, our representation is indeed stable under changes to these scales as designed and quickly converges to consistent behaviour as the network size and number of time steps increases.

\begin{figure}
    \centering
    \begin{subfigure}{0.45\textwidth}
    \includegraphics[width=\linewidth]{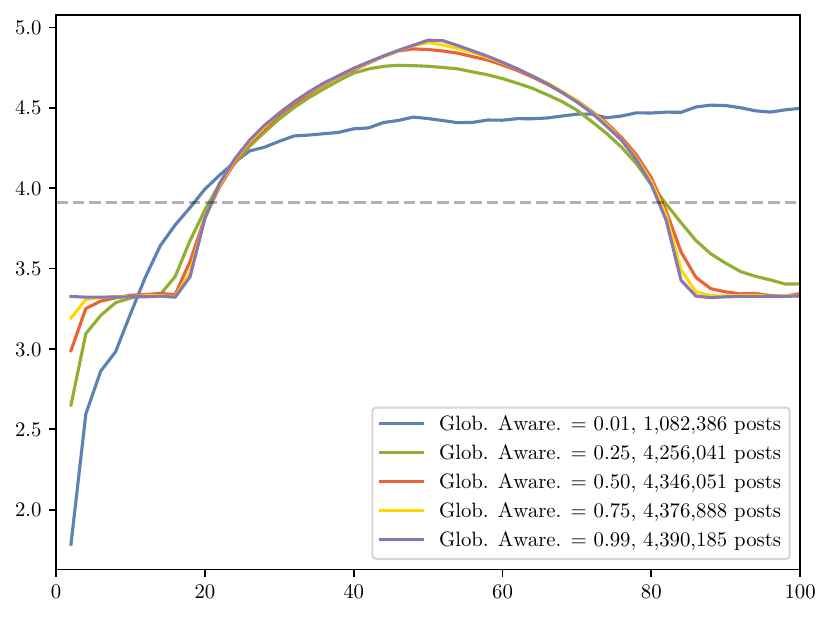}
    \caption{Varying global awareness, determining the likelihood of choosing from locally or globally important topics.}
    \label{fig:vary_ga}
    \end{subfigure}
    \hfill
    \begin{subfigure}{0.45\textwidth}
    \includegraphics[width=\linewidth]{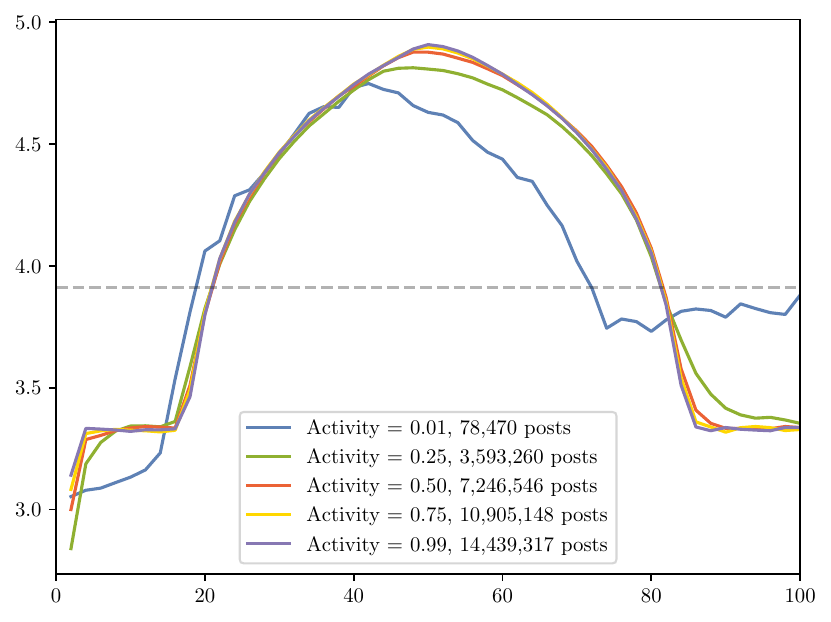}
    \caption{Varying agent activity, determining the likelihood of an agent choosing to post at each time step.}
    \label{fig:vary_act}
    \end{subfigure}
    \\
    \begin{subfigure}{0.45\textwidth}
    \includegraphics[width=\linewidth]{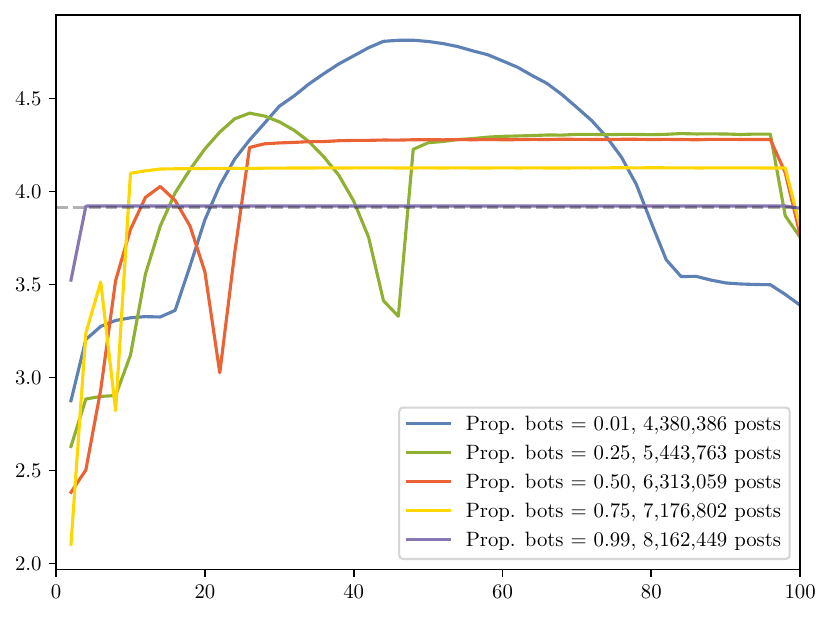}
    \caption{Varying the number of spamming agents. Each such agent activated between time steps 375 and 475.}
    \label{fig:vary_bots}
    \end{subfigure}
    \hfill
    \begin{subfigure}{0.45\textwidth}
    \includegraphics[width=\linewidth]{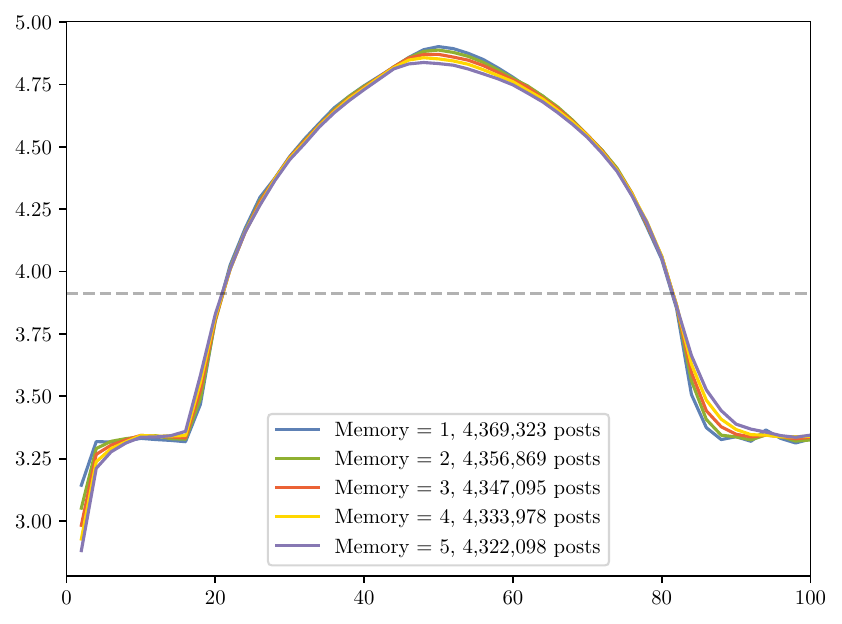}
    \caption{Varying agent memory, determining the number of previous posts considered for local awareness.}
    \label{fig:vary_mem}
    \end{subfigure}
    \caption{The impact of running our agent-based model with different agent settings. The dashed line indicates the value of uniform, constant behaviour across the simulation.}
    \label{fig:vary_agent_props}
\end{figure}

The first parameter we choose to vary is the global awareness, determining the likelihood that an agent will choose globally relevant topics rather than locally relevant topics. Fig.~\ref{fig:vary_ga} shows our scale-independent representation of the model posts. Varying the global awareness parameter reveals two trends. Under low global awareness, we see a small injection of content during the event period before local sharing takes over and maintains a stable level of activity for the remainder of the simulation. This local sharing sees limited total uptake across the system however and reports significantly lower total on-topic activity than the other cases. For levels of global attention above 25\% however the qualitative trends remain largely unchanged wherein the event period drives most of the on-topic activity and should therefore be considered as the expected behaviour under the default parameters. This observation is reinforced by the total number of topical posts made, which grows slowly after this point.

In Fig.~\ref{fig:vary_act} we show the simulation results for varying agent activity, that is the likelihood that an agent will post at a given time step. As expected this parameter has a significant impact on the total number of topical posts made during the simulation. We do see however that the relative posting frequency of agents has limited impact on the behaviour demonstrated under the scale-independent representation and beyond modest individual activity rates we approach a steady state with qualitatively similar behaviour for a wide range of activity rates. In the experiments shown in Fig.~\ref{fig:vary_act} we once again see that the rate of activity returns to that seen in the pre-event period. This figure also highlights the event focus characteristic of our scale-independent representation. In the case of very low activity, we see increased persistence in the topic after the peak in global attention when compared to higher activity rates. This is due to the lower activity rate reducing the difference between the number of on-topic posts made in each of these two periods.

We now vary the proportion of spamming users in the system and illustrate the behaviour in Fig.~\ref{fig:vary_bots}. The simulations we consider here have a single period of spamming activity covering one fifth of the total simulation and occurring after the global attention function has returned to its baseline. Even with relatively small proportions of spamming users in the population, we can see that the total number of posts using the topic of interest increases. Moreover, the relative size of the periods of authentic and spamming behaviour are quickly shifted in favour of the spamming behaviour. We once again see here how the scale-independent representation focuses in on periods of increased activity. Consequently, as the total number of posts in the spamming period increases to cover larger proportions of the total number of posts, this leads to larger periods of the stable activity we argue is characteristic of the abrupt-shift shape. Of particular note is the case where 99\% of all users follow this spamming behaviour - here we see that behaviour is approximately equivalent to a steady rate across most of the total activity. One other aspect that our simulation reveals in comparison with the Twitter example is the behaviour during the spamming period. Our spamming agents post consistently and regularly and therefore produce stable behaviour. By contrasting this with the examples from the Twitter data, the inherent variability in background Twitter usage rates or the activity of parties attempting to manipulate the conversation is likely to be more variable (relative to the typical user patterns) than those included in our simulations.

Our final exploration of the impact of agent parameters on the type of behaviour seen concerns the number of posts from each neighbour an agent considers when using local preferences. In Fig.~\ref{fig:vary_mem} we vary the total number of posts to consider. Here we see that the memory parameter does not qualitatively affect the type of behaviour seen in the simulation. In all test cases, the activity profile is driven by the change in global attention and returns to a baseline rate after the event period. There is an inverse correlation between memory size and total topical posts, but the impact is small relative to the total post count.

Considering these simulation results together, we observe that varying properties of the agents has limited impact on the behaviour produced by the model. The exceptions are in the case of low engagement with the system at large (expressed in global awareness and activity). This stands in contrast to the proportion of spamming agents in the simulation, which quickly dominates authentic activity. As a result, we are confident that our chosen set of default model parameters are representative of typical activity in a range of settings.

\pagebreak
\section{Applying interval division for further analysis of \#cpc18}\label{sec:splitting-cpc18}

\begin{figure}[h]
    \centering
    \includegraphics[width=0.5\linewidth]{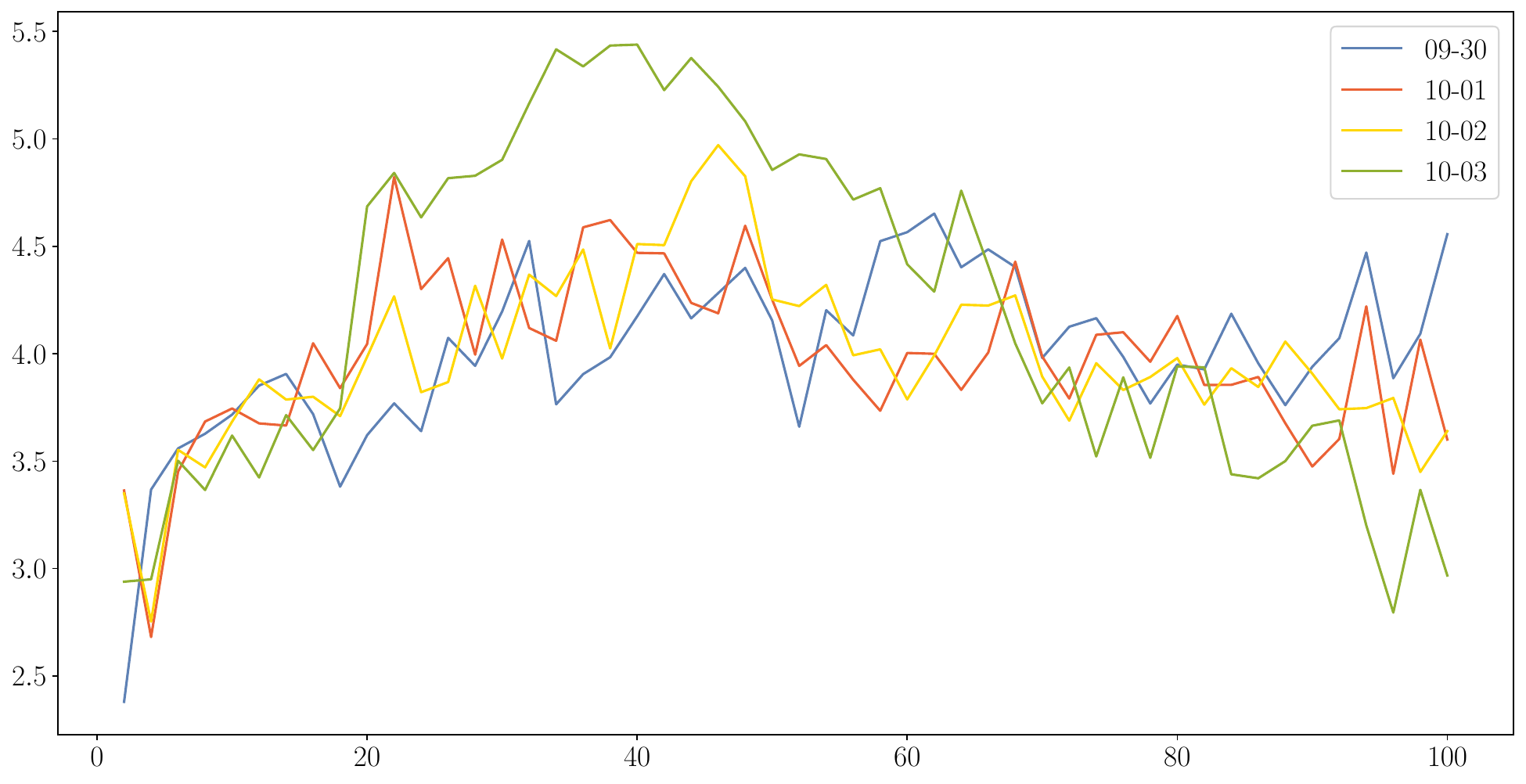}
    \caption{Division of \#cpc18 into days. We compute the scale-independent representation for each of the four conference days individually and find that each is characterised by some growth in attention during the day, with persistent activity for each of the first three days. The final day (10-03) reveals an arch-shaped profile.}
    \label{fig:cpc18-split}
\end{figure}

The example \#cpc18 shown in Fig.~\ref{sh:fig:cpc18} shows valuable utility for the scale-independent representation in the comparison of multiple events occurring within a short time frame. In Fig.~\ref{sh:fig:cpc18_d} we consider the entire period of the conference as a single interval. Here we see sudden increases and decreases before and after the conference, but periodic behaviour suggests that this event is not an example of abrupt shift behaviour. In Fig.~\ref{fig:cpc18-split} we focus instead on the four conference days and visualise the scale-independent representation for each in turn. Here we can more clearly see the patterns alluded to by Fig.~\ref{sh:fig:cpc18_d}, that this interval contains several individual events. Given their close proximity their is insufficient time for attention to return to its baseline but the scale-independent representation is sufficient to reveal this pattern without additional intervention.

\end{document}